\documentclass[prd,showkeys,floatfix,twocolumn,amsmath,amssymb,floatfix]{revtex4}
\usepackage{graphicx,color,dcolumn,booktabs,bm}
\usepackage{longtable,lscape}
\usepackage{amssymb}
\usepackage{indentfirst}
\usepackage{epsfig}
\usepackage{feynmf}
\usepackage{epstopdf}
\usepackage{slashed}
\usepackage{cases}
\usepackage{multirow}
\usepackage[colorlinks,citecolor=blue,anchorcolor=red,menucolor=red,linkcolor=red,filecolor=red,runcolor=red,urlcolor=blue,frenchlinks=red]{hyperref}

\newcommand{\feynp}[1]{#1\kern-0.45em/}
\def\dab{\int^{\alpha_{max}}_{\alpha_{min}}d\alpha\int^{\beta_{max}}_{\beta_{min}}d\beta}
\def\qq{\langle\bar qq\rangle}

\def\GGa{\langle GG\rangle}
\def\GGb{\langle g_s^2GG\rangle}

\def\qGqa{\langle\bar qGq\rangle}

\def\qGqb{\langle g_s\bar q\sigma Gq\rangle}

\def\FF(s){\left[(\alpha+\beta)m_c^2-\alpha\beta s\right]}
\def\HH(s){\left[m_c^2-\alpha(1-\alpha) s\right]}
\def\KK(s){\left[\gamma m_c^2-\gamma(1-\gamma) s\right]}
\def\non{\\ \nonumber}

\allowdisplaybreaks[3]

\begin{document}
\title{Relative $P$-wave excitations of $D^{(*)}\bar D^{(*)}$ configurations in QCD sum rules}
%

\author{Yi-Wei Jiang$^1$}
\author{Hua-Xing Chen$^1$}
\email{hxchen@seu.edu.cn}
\author{Er-Liang Cui$^2$}
\email{erliang.cui@nwafu.edu.cn}
\author{Ding-Kun Lian$^1$}
\email{liandk@seu.edu.cn}
\author{Niu Su$^3$}
\email{suniu@tyut.edu.cn}

\affiliation{
$^1$School of Physics, Southeast University, Nanjing 210094, China
\\
$^2$College of Science, Northwest A\&F University, Yangling 712100, China
\\
$^3$College of Physics and Optoelectronic Engineering, Taiyuan University of Technology, Taiyuan 030024, China
}

\begin{abstract}

We investigate the relative $P$-wave excitations of the
$D^{(*)}\bar D^{(*)}$ systems using the QCD sum rule
method. By introducing a covariant derivative between two ground-state
charmed meson fields, we construct color-singlet interpolating currents
to describe the relative orbital excitation between the two meson
clusters. The $P$-wave $D\bar D$, $D\bar D^*/D^*\bar D$, and
$D^*\bar D^*$ configurations with $J^{P}=1^{-}$ are systematically
studied. Their masses are predicted to be around
$3.8$, $3.9$, and $4.0$--$4.1$~GeV, respectively. These values are
close to the corresponding thresholds and are compatible with the
structures around $X(3790)$ and $G(3900)$,
supporting their possible interpretations as $P$-wave
$D\bar D$ and $D\bar D^*/D^*\bar D$ configurations,
respectively. In addition, we find that the $J^{PC}=1^{-+}$ and
$1^{--}$ combinations in the $D\bar D^*/D^*\bar D$ sector possess
identical correlation functions in the present QCD sum rule framework,
leading to degenerate mass predictions for these two channels. These results indicate that relative $P$-wave excitations of heavy
meson pairs may constitute another interesting class of near-threshold
hadronic configurations worthy of further experimental investigation.

\end{abstract}
\maketitle
\pagenumbering{arabic}

\section{Introduction}
\label{sec:intro}

During the past two decades, many candidates for exotic hadrons have
been observed in particle experiments, providing an important platform
for studying the nonperturbative dynamics of the strong interaction~\cite{pdg}. Among the various theoretical interpretations, the
hadronic molecular picture is particularly attractive for states located
close to two-hadron thresholds. A well-known example is the
$X(3872)$, whose mass is extremely close to the $D^0\bar D^{*0}$
threshold~\cite{Belle:2003nnu}. Similar near-threshold structures
have subsequently been observed in both hidden-charm and doubly-charm
systems, stimulating extensive studies of the interactions among charmed
mesons~\cite{Chen:2016qju,Liu:2019zoy,Chen:2022asf,Luo:2025sns,
Hosaka:2016pey,Richard:2016eis,Lebed:2016hpi,Esposito:2016noz,
Ali:2017jda,Guo:2017jvc,Olsen:2017bmm,Karliner:2017qhf,
Guo:2019twa,Brambilla:2019esw,Yang:2020atz,Fang:2021wes,
Jin:2021vct,Meng:2022ozq,Liu:2024uxn,Wang:2025dur,Dai:2026fkg,Bai:2026atm}.

Most previous studies of hadronic molecules have concentrated on
$S$-wave systems, while their orbital excitations have received
considerably less attention. Nevertheless, the relative $P$-wave
interactions among $D\bar D$, $D\bar D^*/D^*\bar D$, and
$D^*\bar D^*$ mesons may also generate interesting near-threshold
structures. The $P$-wave coupled-channel dynamics involving these
thresholds were systematically investigated in
Ref.~\cite{Du:2016qcr}, where possible $P$-wave molecular
structures were explored and a pole close to the
$D\bar D^*/D^*\bar D$ threshold was found. More recently, the BESIII
Collaboration performed a precise measurement of the
$e^+e^-\to D\bar D$ Born cross sections and observed an enhancement
around $3.9$~GeV~\cite{BESIII:2024ths}. This structure was
subsequently proposed in Ref.~\cite{Lin:2024qcq} as a possible
$P$-wave $D\bar D^*/D^*\bar D$ molecular resonance. These studies
suggest that $P$-wave charmed-meson systems may provide another
interesting class of exotic hadrons beyond the extensively studied
$S$-wave molecules. We refer to Refs.~\cite{Huang:2025rvj,Cao:2025okk,Lu:2025zae,Chen:2025gxe,Ye:2025ywy,Liu:2025sjz,Chen:2025mgg,Qian:2025zyp,Wang:2026mzg,Lu:2026zcs,Abreu:2026ngb} for more relevant discussions.

Very recently, the BESIII Collaboration reported the observation of a
charged charmoniumlike structure, denoted as $X(3790)^-$, in the
$D^0D^-$ invariant-mass spectrum of the
$e^+e^-\to D^0D^-\pi^+$ process~\cite{BESIII:2026emz}.
Its significance was determined to be $5.3\sigma$. Its quantum number was
determined to be $J^P=1^-$, and its mass was measured to be
$3794.7\pm6.0\pm8.7$~MeV. Since two ground-state $D$ mesons can form
a $J^P=1^-$ configuration through a relative $P$-wave, the
$X(3790)^-$ provides a particularly interesting motivation for a
systematic investigation of the
$P$-wave $D^{(*)}\bar D^{(*)}$ meson-meson systems.

The method of QCD sum rules has been extensively applied to study
multiquark and hadronic molecular states. Most molecular QCD sum rule
calculations employ local color-singlet meson-meson currents, which
mainly describe $S$-wave configurations without explicitly incorporating
the relative orbital excitation between the two meson clusters
~\cite{Cui:2013yva,Chen:2013omd,Wang:2013daa}.
On the other hand, derivative interpolating currents have been widely
used in QCD sum rule studies to describe orbital excitations of
conventional hadrons and compact multiquark states
~\cite{Chen:2015kpa,Su:2022eun,Wang:2018ejf,Xin:2022qnv,Xi:2024cvy}. Derivative operators have also
appeared in molecular-current constructions, for example in the
$D^*\bar D^*$ QCD sum rule study of
Ref.~\cite{Cui:2013xla}. However, to the best of our knowledge,
the relative $P$-wave $D^{(*)}\bar D^{(*)}$ systems have not yet been
systematically investigated using color-singlet meson-meson currents
with a covariant derivative explicitly describing the relative orbital
motion between two ground-state charmed mesons.

\begin{figure*}[t]
\begin{center}
\includegraphics[width=0.96\textwidth]{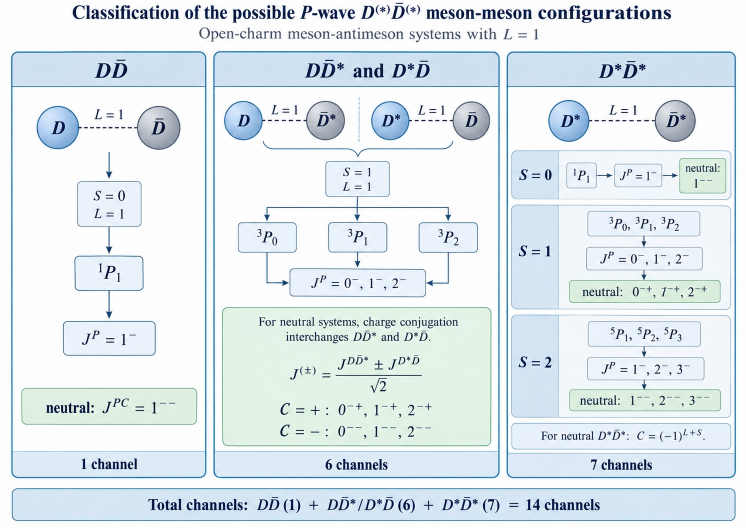}
\caption{
Classification of the possible $P$-wave
$D^{(*)}\bar D^{(*)}$ meson-meson configurations.
The $D\bar D$, $D\bar D^*/D^*\bar D$, and
$D^*\bar D^*$ sectors contain one, six, and seven channels,
respectively.
For the neutral configurations, the corresponding $C$ parities are
also indicated.
}
\label{fig:classification}
\end{center}
\end{figure*}

In this paper, we apply the method of QCD sum rules to systematically
study the relative $P$-wave $D^{(*)}\bar D^{(*)}$ meson-meson systems.
We construct the relevant interpolating currents in the generic form
\begin{equation}
J_\lambda
\sim
[\bar q\Gamma_1 c]\,
\overleftrightarrow D_\lambda\,
[\bar c\Gamma_2 q] ,
\end{equation}
where the covariant derivative explicitly introduces the relative
$P$-wave excitation between the two color-singlet meson clusters.
We investigate the $D\bar D$,
$D\bar D^*/D^*\bar D$, and $D^*\bar D^*$ configurations and classify
their possible quantum numbers. Their two-point correlation functions
are calculated using the operator product expansion, and the
corresponding mass spectra are extracted through QCD sum rule analyses.
Particular attention is paid to the $J^P=1^-$ channels, which may be
related to the recently observed $X(3790)^-$ in the $D\bar D$ channel
and the $G(3900)$ structure in the $D\bar D^*/D^*\bar D$ channel.

This paper is organized as follows. In Sec.~\ref{sec:current}, we
systematically construct and classify the interpolating currents for
the $P$-wave $D^{(*)}\bar D^{(*)}$ systems. In
Sec.~\ref{sec:sumrule}, we derive the corresponding QCD sum rules and
calculate the operator product expansion. Numerical analyses are
performed in Sec.~\ref{sec:numerical}. The obtained results are
summarized and discussed in Sec.~\ref{sec:summary}.

\section{Interpolating Currents}
\label{sec:current}

In this section, we systematically construct the interpolating currents
for the $P$-wave $D^{(*)}\bar D^{(*)}$ meson-meson configurations.
We use $c$ to denote the charm quark and $q=u,d$ to denote a light quark.
The following color-singlet quark-antiquark operators are used to
interpolate the $D$ and $D^*$ mesons:
\begin{eqnarray}
j^D &=& \bar q_a \gamma_5 c_a \, ,
\\
j^{D^*}_\mu &=& \bar q_a \gamma_\mu c_a \, ,
\\
j^{\bar D} &=& \bar c_a \gamma_5 q_a \, ,
\\
j^{\bar D^*}_\mu &=& \bar c_a \gamma_\mu q_a \, ,
\end{eqnarray}
where $a$ is the color index.

To describe the relative $P$-wave excitation between the two mesonic
clusters, we introduce one covariant derivative into the local
meson-meson interpolating currents:
\begin{equation}
\big[ A \overset{\leftrightarrow}{D}_\lambda B \big]
\equiv
A[D_\lambda B]-[D_\lambda A]B \, .
\label{def:derivative}
\end{equation}
The covariant derivative acting on the constituent quark fields is
defined as
\begin{equation}
D_\lambda
=
\partial_\lambda
+
i g_s A_\lambda^A t^A \, ,
\end{equation}
with the corresponding opposite sign for its action on an antiquark
field. In the QCD sum rule calculations, the derivative in
Eq.~(\ref{def:derivative}) is implemented explicitly at the
quark-field level. Consequently, both the ordinary derivative terms
and the gluonic terms proportional to $g_s A_\lambda^A t^A$ are
included.

Accordingly, we introduce the following four basic $P$-wave
meson-meson operators, where the light-flavor labels are suppressed
for simplicity:
\begin{eqnarray}
J^{D\bar D}_{\lambda}
&=&
(\bar q_a\gamma_5c_a)
\overset{\leftrightarrow}{D}_{\lambda}
(\bar c_b\gamma_5q_b) \, ,
\\
J^{D\bar D^*}_{\lambda\mu}
&=&
(\bar q_a\gamma_5c_a)
\overset{\leftrightarrow}{D}_{\lambda}
(\bar c_b\gamma_\mu q_b) \, ,
\\
J^{D^*\bar D}_{\lambda\mu}
&=&
(\bar q_a\gamma_\mu c_a)
\overset{\leftrightarrow}{D}_{\lambda}
(\bar c_b\gamma_5q_b) \, ,
\\
J^{D^*\bar D^*}_{\lambda\mu\nu}
&=&
(\bar q_a\gamma_\mu c_a)
\overset{\leftrightarrow}{D}_{\lambda}
(\bar c_b\gamma_\nu q_b) \, .
\end{eqnarray}
Here the Lorentz index $\lambda$ is associated with the relative
$P$-wave excitation, while $\mu$ and $\nu$ are associated with
the spin-one $D^*$ and $\bar D^*$ mesons, respectively.
After coupling the orbital and spin indices to definite total angular
momentum, we use $\alpha$, $\beta$, and $\gamma$ to denote the free
Lorentz indices of the resulting projected currents.
Different couplings among $\lambda$, $\mu$, and $\nu$ lead to
configurations with different total angular momenta, which will be
systematically constructed below. The overall classification is
summarized in Fig.~\ref{fig:classification}.

\subsection{$D\bar D$ configuration}

Both $D$ and $\bar D$ mesons have spin zero, so that their total
spin is $S=0$. Combining it with the relative orbital angular
momentum $L=1$, we obtain only one $P$-wave configuration,
\begin{equation}
{}^1P_1:\qquad J^P=1^- \, .
\end{equation}
For the neutral $D\bar D$ system, this configuration has $J^{PC}=1^{--}$. The corresponding interpolating current is obtained directly from
$J^{D\bar D}_{\lambda}$ as
\begin{equation}
J^{D\bar D[1^{--}]}_{\alpha}
=
g_{\alpha\lambda} J^{D\bar D;\lambda}
=
(\bar q_a\gamma_5c_a)
\overset{\leftrightarrow}{D}_{\alpha}
(\bar c_b\gamma_5q_b) \, .
\label{current:DD1}
\end{equation}
For the neutral system, this vector current has negative
charge parity and can in general couple to both spin-one and
spin-zero states through
\begin{eqnarray}
\langle 0 |
J^{D\bar D[1^{--}]}_{\alpha}
| D\bar D;1^{--}(q,\epsilon) \rangle
&=&
f_{1^{--}}\epsilon_\alpha \, ,
\label{coupling:DD1}
\\
\langle 0 |
J^{D\bar D[1^{--}]}_{\alpha}
| D\bar D;0^{+-}(q) \rangle
&=&
f_{0^{+-}}q_\alpha \, .
\label{coupling:DD0}
\end{eqnarray}
The latter longitudinal component therefore carries the exotic
quantum numbers $J^{PC}=0^{+-}$.

For later use, we introduce the transverse tensor
\begin{equation}
\widetilde g_{\alpha\beta}
=
g_{\alpha\beta}
-
\frac{q_\alpha q_\beta}{q^2} \, .
\label{def:gtilde}
\end{equation}
The corresponding two-point correlation function can be decomposed as
\begin{eqnarray}
&& \nonumber\Pi_{\alpha\beta}(q)
\\ \nonumber
&\equiv&
i\int d^4x\,e^{iqx}
\langle0|
{\bf T}
\left[
J^{D\bar D[1^{--}]}_{\alpha}(x)
J^{D\bar D[1^{--}],\dagger}_{\beta}(0)
\right]
|0\rangle
\nonumber\\
&=&
-\widetilde g_{\alpha\beta}\Pi_{1}(q^2)
+
\frac{q_\alpha q_\beta}{q^2}
\Pi_{0}(q^2) \, .
\label{correlator:DD}
\end{eqnarray}
The $J^{PC}=1^{--}$ component of interest can therefore be isolated
through the transverse Lorentz structure
$\widetilde g_{\alpha\beta}$, or equivalently through the
$g_{\alpha\beta}$ Lorentz structure of the correlation function.

\subsection{$D\bar D^*$ and $D^*\bar D$ configurations}

For the $D\bar D^*$ and $D^*\bar D$ systems, the total spin of the
two mesons is $S=1$. Combining it with the relative orbital angular momentum $L=1$, we have $1\otimes1=0\oplus1\oplus2$, corresponding to the three $P$-wave configurations
\begin{equation}
{}^3P_0,\qquad {}^3P_1,\qquad {}^3P_2 \, ,
\end{equation}
with
\begin{equation}
J^P=0^-,\qquad1^-,\qquad2^- \, ,
\end{equation}
respectively.

The three projected $D\bar D^*$ currents are constructed from
the basic operator $J^{D\bar D^*}_{\lambda\mu}$ as
\begin{eqnarray}
J^{D\bar D^*[0^-]}
&=&
g_{\lambda\mu}
J^{D\bar D^*;\lambda\mu} \, ,
\label{current:DDs0}
\\
J^{D\bar D^*[1^-]}_{\alpha\beta}
&=&
{1\over2}
\left(
g_{\alpha\lambda}g_{\beta\mu}
-
g_{\beta\lambda}g_{\alpha\mu}
\right)
J^{D\bar D^*;\lambda\mu} \, ,
\label{current:DDs1}
\\
J^{D\bar D^*[2^-]}_{\alpha\beta}
&=&
{\cal P}^{(2)}_{\alpha\beta;\lambda\mu}
J^{D\bar D^*;\lambda\mu} \, ,
\label{current:DDs2}
\end{eqnarray}
where the four-dimensional symmetric-traceless rank-two projector
is defined as
\begin{equation}
{\cal P}^{(2)}_{\alpha\beta;\lambda\mu}
=
{1\over2}
\left(
g_{\alpha\lambda}g_{\beta\mu}
+
g_{\alpha\mu}g_{\beta\lambda}
\right)
-
{1\over4}g_{\alpha\beta}g_{\lambda\mu} \, .
\label{def:P2}
\end{equation}
The corresponding $D^*\bar D$ currents,
$J^{D^*\bar D[J^-]}$ with $J=0,1,2$, are obtained by the replacement
\begin{equation}
J^{D\bar D^*}_{\lambda\mu}
\longrightarrow
J^{D^*\bar D}_{\lambda\mu}
\end{equation}
in Eqs.~(\ref{current:DDs0})--(\ref{current:DDs2}).

For the neutral systems, charge conjugation interchanges the
$D\bar D^*$ and $D^*\bar D$ configurations. With the definitions
adopted above, we have
\begin{equation}
{\mathbb C}
J^{D\bar D^*[J^-]}
{\mathbb C}^{-1}
=
J^{D^*\bar D[J^-]} \, ,
\quad J=0,1,2 \, ,
\end{equation}
and vice versa, where the free Lorentz indices are omitted for
the $J=1$ and $J=2$ currents. We can therefore construct the
$C$-parity eigenstates as
\begin{eqnarray}
J^{D\bar D^*[J^{-+}]}
&=&
{1\over\sqrt{2}}
\left(
J^{D\bar D^*[J^-]}
+
J^{D^*\bar D[J^-]}
\right) \, ,
\\
J^{D\bar D^*[J^{--}]}
&=&
{1\over\sqrt{2}}
\left(
J^{D\bar D^*[J^-]}
-
J^{D^*\bar D[J^-]}
\right) \, ,
\label{current:Cpm}
\end{eqnarray}
which satisfy
\begin{eqnarray}
{\mathbb C}
J^{D\bar D^*[J^{-+}]}
{\mathbb C}^{-1}
&=&
+J^{D\bar D^*[J^{-+}]} \, ,
\\
{\mathbb C}
J^{D\bar D^*[J^{--}]}
{\mathbb C}^{-1}
&=&
-J^{D\bar D^*[J^{--}]} \, .
\end{eqnarray}
Consequently, the $C=+$ currents have $J^{PC}=0^{-+},1^{-+},2^{-+}$, while the $C=-$ currents have $J^{PC}=0^{--},1^{--},2^{--}$. For the charged systems, $C$ parity is not defined, and the
$D\bar D^*$ and $D^*\bar D$ configurations are treated directly
in terms of their $J^P=0^-,1^-,2^-$ currents.

The desired $J^{PC}=0^{-\pm}$, $1^{-\pm}$, and $2^{-\pm}$
components couple to the above currents through
\begin{eqnarray}
\langle 0 |
J^{D\bar D^*[0^{-\pm}]}
| D\bar D^*;0^{-\pm}(q)\rangle
&=&
f_{0^{-\pm}} \, ,
\label{coupling:DDs0}
\\
\langle 0 |
J^{D\bar D^*[1^{-\pm}]}_{\alpha\beta}
| D\bar D^*;1^{-\pm}(q,\epsilon)\rangle
\!&=&\!
f_{1^{-\pm}}
\varepsilon_{\alpha\beta\rho\sigma}
q^\rho\epsilon^\sigma \, ,
\label{coupling:DDs1}
\\
\langle 0 |
J^{D\bar D^*[2^{-\pm}]}_{\alpha\beta}
| D\bar D^*;2^{-\pm}(q,\epsilon)\rangle
&=&
f_{2^{-\pm}}\epsilon_{\alpha\beta} \, .
\label{coupling:DDs2}
\end{eqnarray}
The tensor currents may also couple to states with other spin-parity
quantum numbers but the same charge parity. These contributions can
be separated by choosing appropriate Lorentz structures of the
two-point correlation functions. For this purpose, we use the transverse tensor
$\widetilde g_{\rho\sigma}$ defined in Eq.~(\ref{def:gtilde}).
No further Lorentz decomposition is needed for the scalar
$J^{D\bar D^*[0^{-\pm}]}$ currents.

For the antisymmetric rank-two currents
$J^{D\bar D^*[1^{-\pm}]}_{\alpha\beta}$, the desired
$J^{PC}=1^{-\pm}$ component can be isolated through
\begin{equation}
\widetilde g_{\alpha\gamma}
\widetilde g_{\beta\delta}
-
\widetilde g_{\alpha\delta}
\widetilde g_{\beta\gamma} \, .
\label{structure:DDs1}
\end{equation}
Accordingly, we work directly with the coefficient of this Lorentz
structure and write
\begin{equation}
\Pi^{1^{-\pm}}_{\alpha\beta,\gamma\delta}(q)
=
\left(
\widetilde g_{\alpha\gamma}
\widetilde g_{\beta\delta}
-
\widetilde g_{\alpha\delta}
\widetilde g_{\beta\gamma}
\right)
\Pi_{1^{-\pm}}(q^2)
+\cdots \, ,
\label{correlator:DDs1}
\end{equation}
where the ellipsis denotes Lorentz structures associated with
other spin-parity components.

For the symmetric-traceless rank-two currents
$J^{D\bar D^*[2^{-\pm}]}_{\alpha\beta}$, the desired
$J^{PC}=2^{-\pm}$ component can be isolated through the physical
spin-two projector
\begin{eqnarray}
{\cal P}^{(2)}_{\alpha\beta;\gamma\delta}(q)
&=&
{1\over2}
\left(
\widetilde g_{\alpha\gamma}
\widetilde g_{\beta\delta}
+
\widetilde g_{\alpha\delta}
\widetilde g_{\beta\gamma}
\right)
-
{1\over3}
\widetilde g_{\alpha\beta}
\widetilde g_{\gamma\delta} \, ,
\label{def:P2physical}
\end{eqnarray}
so that
\begin{equation}
\Pi^{2^{-\pm}}_{\alpha\beta,\gamma\delta}(q)
=
{\cal P}^{(2)}_{\alpha\beta;\gamma\delta}(q)
\Pi_{2^{-\pm}}(q^2)
+\cdots \, .
\label{correlator:DDs2}
\end{equation}
%

\subsection{$D^*\bar D^*$ configuration}

The spins of the two vector mesons can couple to $1\otimes1=0\oplus1\oplus2$, corresponding to the total spins $S=0$, $1$, and $2$, respectively.
Combining these spin configurations with the relative orbital
angular momentum $L=1$, the $D^*\bar D^*$ system contains
\begin{eqnarray}
S=0 &: & {}^1P_1 \, ,
\\
S=1 &: & {}^3P_0,\ {}^3P_1,\ {}^3P_2 \, ,
\\
S=2 &: & {}^5P_1,\ {}^5P_2,\ {}^5P_3 \, .
\end{eqnarray}
Altogether, there are seven independent $P$-wave
$D^*\bar D^*$ configurations.

For a neutral $D^*\bar D^*$ system, the charge parity is $C=(-1)^{L+S}$. Since $L=1$, the above configurations have
\begin{eqnarray}
S=0 &: &
J^{PC}=1^{--} \, ,
\\
S=1 &: &
J^{PC}=0^{-+},\ 1^{-+},\ 2^{-+} \, ,
\\
S=2 &: &
J^{PC}=1^{--},\ 2^{--},\ 3^{--} \, .
\end{eqnarray}

\subsubsection{$S=0$}

The spin-zero component of the two vector mesons is obtained by
contracting their Lorentz indices. The resulting $1^{--}$ current is
\begin{eqnarray}
J^{D^*\bar D^*[1^{--\prime}]}_{\alpha}
&=&
g_{\alpha\lambda}g_{\mu\nu}
J^{D^*\bar D^*;\lambda\mu\nu} \, .
\label{current:DsDsS0}
\end{eqnarray}
Here the prime is introduced to distinguish this $S=0$ current from
the other $1^{--}$ current arising from the $S=2$ configuration
discussed below; it serves only as a label and does not indicate a
radial excitation.

\subsubsection{$S=1$}

We first extract the antisymmetric spin-one component of the two
vector-meson indices:
\begin{equation}
A_{\lambda\mu\nu}
\equiv
{1\over2}
\left(
J^{D^*\bar D^*}_{\lambda\mu\nu}
-
J^{D^*\bar D^*}_{\lambda\nu\mu}
\right) ,
\label{def:A}
\end{equation}
which satisfies $A_{\lambda\mu\nu}=-A_{\lambda\nu\mu}$. The $J^{PC}=0^{-+}$ component is obtained from the totally
antisymmetric part:
\begin{equation}
J^{D^*\bar D^*[0^{-+}]}_{\alpha\beta\gamma}
=
g_{\alpha\lambda}g_{\beta\mu}g_{\gamma\nu}
A^{[\lambda\mu\nu]} \, ,
\label{current:DsDs3P0}
\end{equation}
where
\begin{eqnarray}
&& A_{[\lambda\mu\nu]}
\\ \nonumber 
&&\equiv {1\over6}
\Big(
A_{\lambda\mu\nu}
+A_{\mu\nu\lambda}
+A_{\nu\lambda\mu}
-A_{\lambda\nu\mu}
-A_{\mu\lambda\nu}
-A_{\nu\mu\lambda}
\Big) .
\end{eqnarray}

The $J^{PC}=1^{-+}$ component is represented by the trace vector $V_\nu
\equiv A^\mu_{~\mu\nu}$ through
\begin{equation}
J^{D^*\bar D^*[1^{-+}]}_{\alpha}
=
g_{\alpha\nu}V^\nu \, .
\label{current:DsDs3P1}
\end{equation}

After subtracting both the totally antisymmetric part and the
trace-vector part, we obtain the traceless hook tensor
\begin{equation}
H_{\lambda\mu\nu}
=
A_{\lambda\mu\nu}
-
A_{[\lambda\mu\nu]}
-
{1\over3}
\left(
g_{\lambda\mu}V_\nu
-
g_{\lambda\nu}V_\mu
\right) ,
\label{def:H}
\end{equation}
from which the $J^{PC}=2^{-+}$ current is constructed as
\begin{equation}
J^{D^*\bar D^*[2^{-+}]}_{\alpha\beta\gamma}
=
g_{\alpha\lambda}g_{\beta\mu}g_{\gamma\nu}
H^{\lambda\mu\nu} \, .
\label{current:DsDs3P2}
\end{equation}
The tensor $H_{\lambda\mu\nu}$ satisfies
\begin{eqnarray}
H_{\lambda\mu\nu}
+
H_{\lambda\nu\mu}
&=&
0 \, ,
\\
H_{\lambda\mu\nu}
+
H_{\mu\nu\lambda}
+
H_{\nu\lambda\mu}
&=&
0 \, ,
\\
g^{\lambda\mu}H_{\lambda\mu\nu}
=
g^{\lambda\nu}H_{\lambda\mu\nu}
&=&
0 \, .
\end{eqnarray}

\subsubsection{$S=2$}

The spin-two component of the two vector mesons is obtained using
the four-dimensional symmetric-traceless rank-two projector
defined in Eq.~(\ref{def:P2}):
\begin{equation}
T_{\lambda\mu\nu}
\equiv
{\cal P}^{(2)}_{\mu\nu;\rho\sigma}
J^{D^*\bar D^*;\rho\sigma}_{\lambda} \, .
\label{def:T}
\end{equation}
It satisfies $T_{\lambda\mu\nu}
=
T_{\lambda\nu\mu}$ and
$g^{\mu\nu}T_{\lambda\mu\nu}=0$. The $J^{PC}=1^{--}$ component is represented by the trace vector $W_\nu
\equiv T^\mu_{~\mu\nu}$ through
\begin{equation}
J^{D^*\bar D^*[1^{--}]}_{\alpha}
=
g_{\alpha\nu}W^\nu \, .
\label{current:DsDs5P1}
\end{equation}

The $J^{PC}=2^{--}$ component is obtained from the traceless
hook tensor
\begin{equation}
K_{\lambda\mu\nu}
=
T_{\lambda\mu\nu}
-
T_{\mu\lambda\nu}
-
{1\over3}
\left(
g_{\lambda\nu}W_\mu
-
g_{\mu\nu}W_\lambda
\right) ,
\label{def:K}
\end{equation}
through
\begin{equation}
J^{D^*\bar D^*[2^{--}]}_{\alpha\beta\gamma}
=
g_{\alpha\lambda}g_{\beta\mu}g_{\gamma\nu}
K^{\lambda\mu\nu} \, .
\label{current:DsDs5P2}
\end{equation}
The tensor $K_{\lambda\mu\nu}$ satisfies
\begin{eqnarray}
K_{\lambda\mu\nu}
+
K_{\mu\lambda\nu}
&=&
0 \, ,
\\
K_{\lambda\mu\nu}
+
K_{\mu\nu\lambda}
+
K_{\nu\lambda\mu}
&=&0 \, ,
\\
g^{\lambda\mu}K_{\lambda\mu\nu}
=
g^{\lambda\nu}K_{\lambda\mu\nu}
=
g^{\mu\nu}K_{\lambda\mu\nu}
&=&0 \, .
\end{eqnarray}

Finally, the $J^{PC}=3^{--}$ component is obtained by extracting
the totally symmetric and traceless part of
$T_{\lambda\mu\nu}$. We first define its totally symmetric part as
\begin{equation}
\overline T_{\alpha\beta\gamma}
\equiv
T_{(\alpha\beta\gamma)}
=
{1\over3}
\left(
T_{\alpha\beta\gamma}
+
T_{\beta\gamma\alpha}
+
T_{\gamma\alpha\beta}
\right) ,
\label{def:Tbar}
\end{equation}
where we have used
$T_{\lambda\mu\nu}=T_{\lambda\nu\mu}$.
Its trace is
\begin{equation}
\overline T_\gamma
\equiv
g^{\alpha\beta}\overline T_{\alpha\beta\gamma}
=
{2\over3}W_\gamma \, .
\end{equation}
The $J^{PC}=3^{--}$ current is then constructed as
\begin{equation}
J^{D^*\bar D^*[3^{--}]}_{\alpha\beta\gamma}
=
\overline T_{\alpha\beta\gamma}
-
{1\over6}
\left(
g_{\alpha\beta}\overline T_\gamma
+
g_{\alpha\gamma}\overline T_\beta
+
g_{\beta\gamma}\overline T_\alpha
\right) .
\label{current:DsDs5P3}
\end{equation}
It satisfies $J_{\alpha\beta\gamma}
=
J_{(\alpha\beta\gamma)}$ and 
$g^{\alpha\beta}J_{\alpha\beta\gamma}=0$.

\subsubsection{Couplings and Lorentz structures}

The desired physical components couple to the above currents through
the following matrix elements. For the $S=0$ configuration,
\begin{equation}
\langle0|
J^{D^*\bar D^*[1^{--\prime}]}_\alpha
|D^*\bar D^*;1^{--\prime}(q,\epsilon)\rangle
=
f_{1^{--\prime}}\epsilon_\alpha \, .
\label{coupling:DsDsS0}
\end{equation}
For the $S=1$ configurations,
\begin{eqnarray}
\langle0|
J^{D^*\bar D^*[0^{-+}]}_{\alpha\beta\gamma}
|D^*\bar D^*;0^{-+}(q)\rangle
&=&
f_{0^{-+}}
\varepsilon_{\alpha\beta\gamma\rho}q^\rho \, ,
\\
\langle0|
J^{D^*\bar D^*[1^{-+}]}_\alpha
|D^*\bar D^*;1^{-+}(q,\epsilon)\rangle
&=&
f_{1^{-+}}\epsilon_\alpha \, ,
\\
\langle0|
J^{D^*\bar D^*[2^{-+}]}_{\alpha\beta\gamma}
|D^*\bar D^*;2^{-+}(q,\epsilon)\rangle
&=&
f_{2^{-+}}
\label{coupling:DsDsS1}
\\
\nonumber &&\times
\left(
q_\beta\epsilon_{\alpha\gamma}
-
q_\gamma\epsilon_{\alpha\beta}
\right) .
\end{eqnarray}
For the $S=2$ configurations,
\begin{eqnarray}
\langle0|
J^{D^*\bar D^*[1^{--}]}_\alpha
|D^*\bar D^*;1^{--}(q,\epsilon)\rangle
&=&
f_{1^{--}}\epsilon_\alpha \, ,
\\
\langle0|
J^{D^*\bar D^*[2^{--}]}_{\alpha\beta\gamma}
|D^*\bar D^*;2^{--}(q,\epsilon)\rangle
&=&
f_{2^{--}} \label{coupling:DsDsS2}
\\ 
\nonumber &&\times
\left(
q_\alpha\epsilon_{\beta\gamma}
-
q_\beta\epsilon_{\alpha\gamma}
\right) ,
\\
\langle0|
J^{D^*\bar D^*[3^{--}]}_{\alpha\beta\gamma}
|D^*\bar D^*;3^{--}(q,\epsilon)\rangle
&=&
f_{3^{--}}
\epsilon_{\alpha\beta\gamma} \, .
\end{eqnarray}
As in the previous subsections, the Lorentz currents may also
contain additional physical-spin components. In the QCD sum rule
analysis, the desired states are isolated by selecting the Lorentz
structures generated by the couplings above.

For the vector currents
$J^{D^*\bar D^*[1^{--\prime}]}_\alpha$,
$J^{D^*\bar D^*[1^{-+}]}_\alpha$, and
$J^{D^*\bar D^*[1^{--}]}_\alpha$,
the desired spin-one components are isolated through the transverse
tensor $\widetilde g_{\alpha\beta}$ defined in
Eq.~(\ref{def:gtilde}).
The $J^{PC}=0^{-+}$ component is isolated through the Lorentz
structure generated by
$\varepsilon_{\alpha\beta\gamma\rho}q^\rho$.

For the $J^{PC}=2^{-+}$ and $2^{--}$ currents, the desired spin-two
components are isolated through the hook-type Lorentz structures
generated by Eqs.~(\ref{coupling:DsDsS1}) and
(\ref{coupling:DsDsS2}), together with the physical spin-two
polarization projector ${\cal P}^{(2)}(q)$ defined in
Eq.~(\ref{def:P2physical}).

For the spin-three current, we define the physical transverse
spin-three polarization projector as
\begin{equation}
{\cal P}^{(3)}_{\alpha\beta\gamma;\rho\sigma\tau}(q)
\equiv
\sum_{\rm pol}
\epsilon_{\alpha\beta\gamma}
\epsilon^*_{\rho\sigma\tau} \, ,
\label{def:P3physical}
\end{equation}
where the polarization tensor is totally symmetric, transverse,
and traceless:
\begin{equation}
\epsilon_{\alpha\beta\gamma}
=
\epsilon_{(\alpha\beta\gamma)},
\quad
q^\alpha\epsilon_{\alpha\beta\gamma}=0,
\quad
g^{\alpha\beta}\epsilon_{\alpha\beta\gamma}=0 \, .
\end{equation}
Accordingly, the desired $J^{PC}=3^{--}$ component is extracted from
\begin{equation}
\Pi^{3^{--}}_{\alpha\beta\gamma;\rho\sigma\tau}(q)
=
{\cal P}^{(3)}_{\alpha\beta\gamma;\rho\sigma\tau}(q)
\Pi_{3^{--}}(q^2)
+\cdots \, ,
\end{equation}
where the ellipsis denotes Lorentz structures associated with
other physical-spin components.

\section{QCD Sum Rule Analysis}
\label{sec:sumrule}

In this section, we apply the QCD sum rule method to investigate the
$P$-wave $D^{(*)}\bar D^{(*)}$ meson-meson configurations constructed
in Sec.~\ref{sec:current}. In the present study we concentrate on the
$J^{P}=1^{-}$ channels. Since the corresponding interpolating currents
carry different Lorentz structures, we illustrate the formalism using
the $D\bar D^*/D^*\bar D$ current with $J^{PC}=1^{--}$ as an example.
The same procedure is applied to the other $1^{-}$ currents.

For the current
$J^{D\bar D^*[1^{--}]}_{\alpha\beta}$ defined in
Eq.~(\ref{current:Cpm}), we consider the two-point correlation function
\begin{eqnarray}
&&\Pi_{\alpha\beta,\gamma\delta}(q)
\\
&=& \nonumber
i\int d^4x\,e^{iqx}
\langle 0|
{\bf T}
\left[
J^{D\bar D^*[1^{--}]}_{\alpha\beta}(x)
J^{D\bar D^*[1^{--}],\dagger}_{\gamma\delta}(0)
\right]
|0\rangle .
\label{def:pi}
\end{eqnarray}
As discussed in Sec.~\ref{sec:current}, the desired $J^{PC}=1^{--}$
component can be isolated through the Lorentz structure
\begin{equation}
\widetilde g_{\alpha\gamma}
\widetilde g_{\beta\delta}
-
\widetilde g_{\alpha\delta}
\widetilde g_{\beta\gamma}.
\end{equation}
Accordingly, the correlation function can be written as
\begin{equation}
\Pi_{\alpha\beta,\gamma\delta}(q)
=
\left(
\widetilde g_{\alpha\gamma}
\widetilde g_{\beta\delta}
-
\widetilde g_{\alpha\delta}
\widetilde g_{\beta\gamma}
\right)
\Pi_{1^{--}}(q^2)
+\cdots ,
\label{eq:pi1mm}
\end{equation}
where the ellipsis denotes Lorentz structures associated with other
spin-parity components. In the following, we work directly with the
invariant function $\Pi_{1^{--}}(q^2)$ defined as the coefficient of
this Lorentz structure.

We assume that the current couples to a possible
$D\bar D^*/D^*\bar D$ state with $J^{PC}=1^{--}$,
denoted by $X$, through
\begin{equation}
\langle 0|
J^{D\bar D^*[1^{--}]}_{\alpha\beta}
|X(q,\epsilon)\rangle
=
f_X
\varepsilon_{\alpha\beta\rho\sigma}
q^\rho\epsilon^\sigma ,
\label{eq:defg}
\end{equation}
where $f_X$ is the decay constant and $\epsilon_\mu$ is the
polarization vector.

At the hadron level, the invariant function $\Pi_{1^{--}}(q^2)$ can
be written through the dispersion relation
\begin{equation}
\Pi_{1^{--}}(q^2)
=
\int_{4m_c^2}^{\infty}
\frac{\rho^{\rm phen}(s)}
{s-q^2-i\varepsilon}\,ds ,
\label{eq:hadron}
\end{equation}
where $\rho^{\rm phen}(s)$ is the phenomenological spectral density.
We parameterize it by the lowest-lying pole together with higher states
and continuum contributions:
\begin{equation}
\rho^{\rm phen}(s)
=
f_X^2 M_X^2 \delta(s-M_X^2)
+
\rho^{\rm cont}(s) ,
\label{eq:rho}
\end{equation}
where the extra factor $M_X^2$ originates from the kinematic 
factor $q^2$ generated by the coupling in Eq.~(\ref{eq:defg}), with the
corresponding subtraction term omitted since it vanishes after the
Borel transformation.

At the quark-gluon level, we calculate the correlation function using
the operator product expansion (OPE). After performing the Borel
transformation and adopting the quark-hadron duality approximation for
the continuum contribution above the threshold $s_0$, the
continuum-subtracted OPE expression can be written as
\begin{eqnarray}
&&\Pi_{1^{--}}(s_0,M_B^2)
\\
&\equiv& \nonumber
\int_{4m_c^2}^{s_0}
e^{-s/M_B^2}
\rho^{\rm OPE}(s,M_B^2)\,ds
=
f_X^2 M_X^2 e^{-M_X^2/M_B^2}.
\label{eq_fin}    
\end{eqnarray}
Here $\rho^{\rm OPE}(s,M_B^2)$ denotes the OPE expression entering
the Borel-transformed sum rule; its explicit dependence on $M_B^2$
arises from some higher-dimensional condensate contributions. The
hadron mass and decay constant are then extracted through
\begin{eqnarray}
M_X^2
&=&
\frac{
\dfrac{\partial}{\partial(-1/M_B^2)}
\Pi_{1^{--}}(s_0,M_B^2)
}{
\Pi_{1^{--}}(s_0,M_B^2)
},
\label{eq:LSR}
\\
f_X
&=&
\frac{1}{M_X}
\left[
\Pi_{1^{--}}(s_0,M_B^2)
e^{M_X^2/M_B^2}
\right]^{1/2}.
\label{eq:fx}
\end{eqnarray}
%


\begin{widetext}
The OPE expression obtained from the chosen current $J^{D\bar D^*[1^{--}]}_{\alpha\beta}$ can be written as
	\begin{equation}
		\rho^{\rm OPE}(s) = \rho^{pert}(s) + \rho^{\qq}(s) + \rho^{\GGa}(s)+ \rho^{\qGqa}(s) + \rho^{\qq^2}(s)  + \rho^{\qq\qGqa_0}(s) + {\rho^{\qq\qGqa_2}(s) \over M_B^2} \, ,
		\label{ope:eta1}
	\end{equation}
	where
	\begin{eqnarray}
	\nonumber \rho^{pert}(s) &=& \dab \Bigg\{ { \mathcal{F}(s)^5 } \times \frac{1}{81920 \alpha^4 \beta^4 \pi^6}  \times ( - 5 + 21 \alpha^2 - 16 \alpha^3 + 6 \beta + 6 \alpha \beta - 18 \alpha^2 \beta + 9 \beta^2
	\non
	&-& 12 \alpha \beta^2 - 10 \beta^3) \Bigg\} \, ,
	\non
	\rho^{\qq}(s) &=& {\qq } \dab \Bigg\{ { \mathcal{F}(s)^3 }  \times \frac{m_c}{512 \alpha ^3 \beta ^3 \pi^4} \times ( - \alpha^3 + 9 \alpha \beta - 8 \alpha^2 \beta - \beta^2 - 7 \alpha \beta^2 ) \Bigg\} \, ,
	\non
	\rho^{\GGa}(s) &=& {\GGb }  \int^{\alpha_{max}}_{\alpha_{min}}d\alpha \int^{\beta_{max}}_{\beta_{min}}d\beta \Bigg\{ { \mathcal{F}(s)^3 } \times \frac{1}{98304 \alpha^3 \beta^3 \pi^6} \times ( - 3 \alpha - 6 \alpha^2 + 15  \alpha^3 + 5 \beta -  22 \alpha \beta + 15 \alpha^2 \beta
	\non
	&-& 8 \beta^2 +  33 \alpha \beta^2 + \beta^3 ) + { \mathcal{F}(s)^2 } \times \frac{m_c^2}{98304 \alpha^4 \beta^4 \pi^6} \times ( - 5 \alpha^3 + 21 \alpha^5 - 16 \alpha^6 + 6 \alpha^3 \beta + 6 \alpha^4 \beta - 18 \alpha^5 \beta 
    \non
	&+& 9 \alpha^3 \beta^2 - 12 \alpha^4 \beta^2 - 5 \beta^3 + 21 \alpha^2 \beta^3 - 26 \alpha^3 \beta^3 + 6 \beta^4 + 6 \alpha \beta^4 - 18 \alpha^2 \beta^4 + 9 \beta^5 - 12 \alpha \beta^5 - 10 \beta^6 ) \Bigg\} \, ,
	\non
	\rho^{\qGqa}(s) &=& {\qGqb } \int^{\alpha_{max}}_{\alpha_{min}}d\alpha  \Bigg\{ \int^{\beta_{max}}_{\beta_{min}}d\beta \Bigg[
	{ \mathcal{F}(s)^2 } \times \frac{m_c}{2048 \alpha^2 \beta^3 \pi^4} \times ( - 6 \alpha^3 + 36 \alpha \beta - 15 \alpha^2 \beta + 18 \beta^2 - 93 \alpha \beta^2
    \non
    &-& 6 \beta^3 ) \Bigg] + { \mathcal{H}(s)^2 } \times \frac{m_c}{\pi^4} \times \frac{3 (1-2\alpha)^2}{2048 (1-\alpha)^2\alpha^2}  \Bigg\} \, ,
	\non
	\rho^{\qq^2}(s)&=& {\qq^2 } \int^{\alpha_{max}}_{\alpha_{min}}d\alpha \Bigg\{ { \mathcal{H}(s) } \times \frac{m_c^2}{\pi^2} \times \frac{(1-2\alpha)^2}{64 (1-\alpha)\alpha}  \Bigg\} \, ,
	\non
	\rho^{\qq\qGqa_0}(s)&=& {\qq\qGqb } \int^{\alpha_{max}}_{\alpha_{min}}d\alpha \Bigg\{ \frac{m_c^2}{\pi^2} \times \frac{ 3 - 12 \alpha + 20 \alpha^2 - 12 \alpha^3 }{128 (-1+\alpha)} \Bigg\} \, ,
	\non
	\rho^{\qq\qGqa_2}(s)&=&{\qq\qGqb } \int^{1}_{0} \delta\left( s - {m_c^2 \over (1-\alpha) \alpha} \right) d\alpha  \Bigg\{ \frac{m_c^4}{\pi^2} \times \frac{(1-2\alpha)^2}{128 (-1 + \alpha) \alpha} \Bigg\} \, .
	\end{eqnarray}
\end{widetext}
In the above expressions,
$\mathcal{F}(s)=(\alpha+\beta)m_c^2-\alpha\beta s$ and
$\mathcal{H}(s)=m_c^2-\alpha(1-\alpha)s$, while the integration limits are
\begin{eqnarray}
&&\alpha_{\min,\max}
=
\frac{1\mp\sqrt{1-4m_c^2/s}}{2},
\nonumber\\
&&\beta_{\min}
=
\frac{\alpha m_c^2}{\alpha s-m_c^2},
\qquad
\beta_{\max}=1-\alpha .
\label{eq:integration_limits}
\end{eqnarray}
We carry out the OPE up to dimension eight ($D=8$), including the
perturbative contribution, the quark condensate
$\langle\bar q q\rangle$, the gluon condensate
$\langle g_s^2 GG\rangle$, the quark-gluon mixed condensate
$\langle g_s\bar q\sigma G q\rangle$, and their higher-dimensional
combinations. Terms proportional to the light-quark masses are neglected,
and the vacuum saturation approximation is adopted for the
higher-dimensional condensates. The $D=3$ contribution
$\langle\bar q q\rangle$ and the $D=5$ contribution
$\langle g_s\bar q\sigma G q\rangle$ are both proportional to the charm
quark mass and therefore provide important power corrections to the
correlation function.

The OPE expressions for the remaining $J^{P}=1^{-}$ currents studied
in this work are collected in Appendix~\ref{app:ope}. The corresponding
numerical QCD sum rule analyses are presented in the next section.
It is interesting to note that the $J^{PC}=1^{-+}$ current in the
$D\bar D^*/D^*\bar D$ sector leads to the same invariant correlation
function as its $J^{PC}=1^{--}$ counterpart. To demonstrate this
explicitly, we denote the two $J^P=1^-$ building-block currents defined
in Eq.~(\ref{current:DDs1}) by
\begin{equation}
J_1
=
J^{D\bar D^*[1^-]}_{\alpha\beta},
\qquad
J_2
=
J^{D^*\bar D[1^-]}_{\alpha\beta}.
\end{equation}
In the following, the Lorentz indices are omitted for simplicity.
For the invariant correlation functions relevant to the present sum
rules, we explicitly find
\begin{equation}
\Pi_{11}(q^2)
=
\Pi_{22}(q^2),
\qquad
\Pi_{12}(q^2)
=
\Pi_{21}(q^2)
=
0 .
\label{eq:Piij}
\end{equation}
For the neutral combinations, the corresponding charge-conjugation
eigenstates are
\begin{eqnarray}
J_{+}
&=&
\frac{1}{\sqrt{2}}
\left(
J_1+J_2
\right),
\nonumber\\
J_{-}
&=&
\frac{1}{\sqrt{2}}
\left(
J_1-J_2
\right),
\end{eqnarray}
which have $J^{PC}=1^{-+}$ and $1^{--}$, respectively. Their invariant
correlation functions are therefore
\begin{eqnarray}
\Pi_{1^{-+}}
&=&
\frac{1}{2}
\left(
\Pi_{11}
+\Pi_{22}
+\Pi_{12}
+\Pi_{21}
\right),
\nonumber\\
\Pi_{1^{--}}
&=&
\frac{1}{2}
\left(
\Pi_{11}
+\Pi_{22}
-\Pi_{12}
-\Pi_{21}
\right).
\end{eqnarray}
Using Eq.~(\ref{eq:Piij}), we obtain
\begin{equation}
\Pi_{1^{-+}}(q^2)
=
\Pi_{1^{--}}(q^2).
\label{eq:degenerate}
\end{equation}
Therefore, the $1^{-+}$ and $1^{--}$ $D\bar D^*/D^*\bar D$
channels lead to identical QCD sum rules and are predicted to be
degenerate in mass within the present framework. This degeneracy,
however, concerns only the mass determination from the present
two-point correlation functions. The two states carry opposite charge
parities and may have different long-distance interactions, production
mechanisms, and decay properties, which require separate dynamical
investigations.

\section{Numerical Analyses}
\label{sec:numerical}

In this section, we perform numerical QCD sum rule analyses based on
the OPE expressions obtained in Sec.~\ref{sec:sumrule}. We adopt the
following values for the QCD input parameters
~\cite{Yang:1993bp,Narison:2002hk,Gimenez:2005nt,Jamin:2002ev,
Ioffe:2002be,Ovchinnikov:1988gk,Ellis:1996xc,Narison:2011xe,
Narison:2018dcr,pdg}:
\begin{eqnarray}
\nonumber
\langle g_s^2 GG\rangle
&=&
4\pi\times(0.0635\pm0.0035)\ {\rm GeV}^4 ,
\\
\nonumber
\langle\bar qq\rangle
&=&
-(0.240\pm0.010)^3\ {\rm GeV}^3 ,
\\
\langle g_s\bar q\sigma G q\rangle
&=&
-M_0^2\langle\bar qq\rangle ,
\label{condensates}
\\
\nonumber
M_0^2
&=&
(0.8\pm0.2)\ {\rm GeV}^2 ,
\\
\nonumber
m_c
&=&
1.273\pm0.045\ {\rm GeV} .
\end{eqnarray}
The gluon condensate $\langle g_s^2 GG\rangle$ is not yet precisely
determined, and we adopt the value updated in
Ref.~\cite{Narison:2018dcr}. Numerically, its contribution to the
present sum rules is relatively small.

We use the
$D\bar D^*/D^*\bar D$ current
$J^{D\bar D^*[1^{--}]}_{\alpha\beta}$
as an example to illustrate the numerical analysis. The extracted mass
$M_X$ and decay constant $f_X$ depend on two external parameters,
the continuum threshold $s_0$ and the Borel parameter $M_B^2$. Their
working regions are determined according to three standard criteria:
the convergence of the OPE, the pole contribution, and the stability of
the extracted mass with respect to $M_B^2$ and $s_0$.

\begin{figure}[htbp]
\centering
\includegraphics[width=0.9\linewidth]{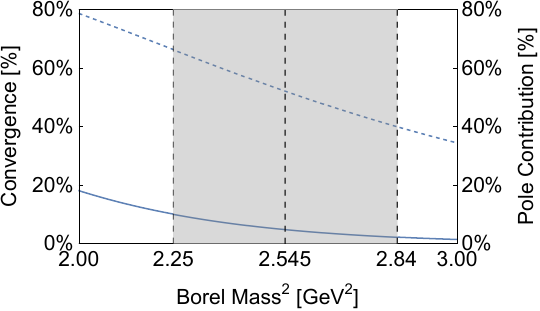}
\caption{
The OPE convergence (CVG, solid curve) and pole contribution
(PC, dashed curve) as functions of the Borel parameter $M_B^2$
for the $D\bar D^*/D^*\bar D$ $J^{PC}=1^{--}$ channel with
$s_0=20.6~{\rm GeV}^2$.
The shaded region,
$2.25~{\rm GeV}^2\leq M_B^2\leq2.84~{\rm GeV}^2$,
denotes the allowed Borel window determined by the requirements
${\rm CVG}\leq10\%$ and ${\rm PC}\geq40\%$.
}
\label{fig:CVGPC}
\end{figure}

We first investigate the convergence of the OPE, which is calculated
up to dimension eight in the present analysis. We require the combined
higher-dimensional $D=6$ and $D=8$ contributions to satisfy
\begin{equation}
{\rm CVG}
\equiv
\left|
\frac{
\Pi_{1^{--}}^{D=6+8}(\infty,M_B^2)
}{
\Pi_{1^{--}}(\infty,M_B^2)
}
\right|
\leq 10\% .
\label{eq:cvg}
\end{equation}
This criterion determines the lower bound of the Borel window. We
further require the pole contribution (PC) to be larger than $40\%$:
\begin{equation}
{\rm PC}
\equiv
\left|
\frac{
\Pi_{1^{--}}(s_0,M_B^2)
}{
\Pi_{1^{--}}(\infty,M_B^2)
}
\right|
\geq 40\% .
\label{eq:pole}
\end{equation}
This requirement determines the upper bound of the Borel window. For
the $J^{PC}=1^{--}$ $D\bar D^*/D^*\bar D$ channel, the CVG and PC
as functions of the Borel parameter $M_B^2$ are shown together in
Fig.~\ref{fig:CVGPC}. Imposing the requirements
${\rm CVG}\leq10\%$ and ${\rm PC}\geq40\%$, we obtain the working
region
\begin{equation}
2.25~{\rm GeV}^2
\leq
M_B^2
\leq
2.84~{\rm GeV}^2 ,
\end{equation}
for $s_0=20.6~{\rm GeV}^2$, as indicated by the shaded region in
Fig.~\ref{fig:CVGPC}.

\begin{figure*}[htbp]
\centering
\includegraphics[width=0.45\linewidth]{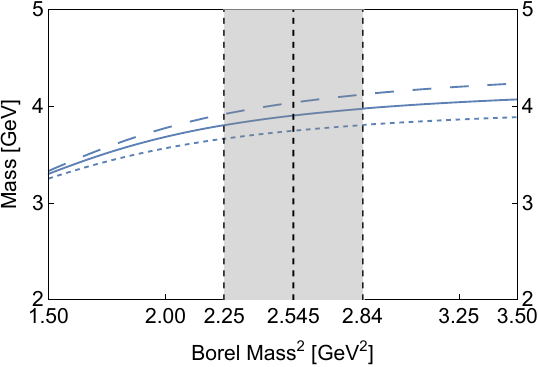}
~~~~~
\includegraphics[width=0.45\linewidth]{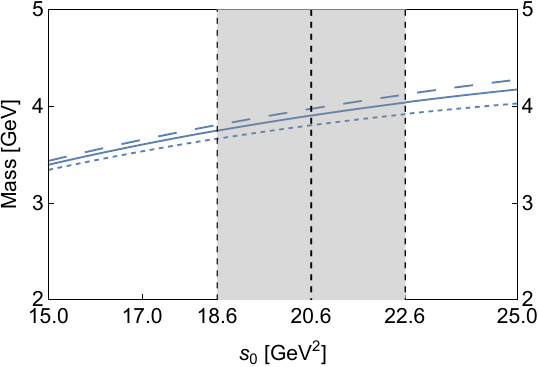}
\caption{
The extracted mass of the
$J^{PC}=1^{--}$ $D\bar D^*/D^*\bar D$ state:
(a) dependence on the Borel parameter $M_B^2$, where the solid curve
corresponds to $s_0=20.6~{\rm GeV}^2$ and the two dashed curves to
$s_0=18.6$ and $22.6~{\rm GeV}^2$;
(b) dependence on the continuum threshold $s_0$, where the solid curve
corresponds to $M_B^2=2.545~{\rm GeV}^2$ and the two dashed curves to
$M_B^2=2.25$ and $2.84~{\rm GeV}^2$.
The shaded regions indicate the working ranges
$2.25\leq M_B^2\leq2.84~{\rm GeV}^2$ and
$18.6\leq s_0\leq22.6~{\rm GeV}^2$, respectively.
}
\label{fig:mass}
\end{figure*}

Finally, within the parameter region allowed by the OPE convergence and
pole dominance criteria, we examine the dependence of the extracted
mass on the Borel parameter and continuum threshold. As shown in
Fig.~\ref{fig:mass}, the extracted mass exhibits a reasonably stable
behavior throughout the working regions of both $M_B^2$ and $s_0$.
Therefore, no further restriction on these parameter regions is imposed. For the present
$J^{PC}=1^{--}$ $D\bar D^*/D^*\bar D$ channel, we take
$s_0=20.6~{\rm GeV}^2$ and $M_B^2=2.50~{\rm GeV}^2$ as the central
parameter values, yielding
\begin{equation}
M_{D\bar D^*[1^{--}]}
=
3.91^{+0.16}_{-0.19}~{\rm GeV}.
\label{eq:mass1}
\end{equation}
The final uncertainty is determined by varying $s_0$ and $M_B^2$
within their working regions and by taking into account the uncertainties
of the QCD input parameters given in Eq.~(\ref{condensates}).

\begin{table*}[hbtp]
\begin{center}
\renewcommand{\arraystretch}{1.5}
\caption{QCD sum rule results for the $P$-wave
$D^{(*)}\bar D^{(*)}$ meson-meson configurations studied in this work.}
\begin{tabular}{>{\hspace{2pt}}c<{\hspace{2pt}}|>{\hspace{2pt}}c<{\hspace{2pt}}|c|c|c|>{\hspace{2pt}}c<{\hspace{2pt}}|>{\hspace{2pt}}c<{\hspace{2pt}}|>{\hspace{2pt}}c<{\hspace{2pt}}}
\hline\hline
\multirow{2}{*}{Current}
&
\multirow{2}{*}{State [$J^{PC},S$]}
&
$s_0^{\rm min}$
&
\multicolumn{2}{c|}{Working Regions}
&
{\rm Pole}
&
Mass
&
Decay Constant
\\
\cline{4-5}
&
&
[{\rm GeV}$^2]$
&
$M_B^2\ [{\rm GeV}^2]$
&
$s_0\ [{\rm GeV}^2]$
&
[\%]
&
[GeV]
&
$f_X$
\\
\hline\hline

$J^{D\bar D[1^{--}]}_{\alpha}$
&
$|D\bar D;1^{--},S=0\rangle$
&
11.8
&
2.00--2.88
&
17.5±2.0
&
40--68
&
$3.77^{+0.13}_{-0.16}$
&
$0.0266^{+0.0103}_{-0.0098}~{\rm GeV}^6$
\\
\hline

$J^{D\bar D^*[1^{--}]}_{\alpha\beta}$
&
$|D\bar D^*/D^*\bar D;1^{--},S=1\rangle$
&
16.2
&
2.25--2.84
&
20.6±2.0
&
40--66
&
$3.91^{+0.16}_{-0.19}$
&
$0.0016^{+0.0006}_{-0.0005}~{\rm GeV}^5$
\\

$J^{D\bar D^*[1^{-+}]}_{\alpha\beta}$
&
$|D\bar D^*/D^*\bar D;1^{-+},S=1\rangle$
&
16.2
&
2.25--2.84
&
20.6±2.0
&
40--66
&
$3.91^{+0.16}_{-0.19}$
&
$0.0016^{+0.0006}_{-0.0005}~{\rm GeV}^5$
\\\hline

$J^{D^*\bar D^*[1^{--\prime}]}_{\alpha}$
&
$|D^*\bar D^*;1^{--},S=0\rangle$
&
19.5
&
2.48--2.84
&
22.5±2.0
&
40--56
&
$4.07^{+0.17}_{-0.20}$
&
$0.0219^{+0.0095}_{-0.0080}~{\rm GeV}^6$
\\

$J^{D^*\bar D^*[1^{-+}]}_{\alpha}$
&
$|D^*\bar D^*;1^{-+},S=1\rangle$
&
13.5
&
1.94--2.74
&
22.5±2.0
&
40--82
&
$4.01^{+0.26}_{-0.37}$
&
$0.0134^{+0.0069}_{-0.0074}~{\rm GeV}^6$
\\

$J^{D^*\bar D^*[1^{--}]}_{\alpha}$
&
$|D^*\bar D^*;1^{--},S=2\rangle$
&
14.4
&
2.07--2.83
&
22.5±2.0
&
40--78
&
$3.99^{+0.23}_{-0.31}$
&
$0.0151^{+0.0070}_{-0.0071}~{\rm GeV}^6$
\\

\hline\hline
\end{tabular}
\label{tab:results}
\end{center}
\end{table*}

\begin{figure}[htbp]
\centering
\includegraphics[width=0.9\linewidth]{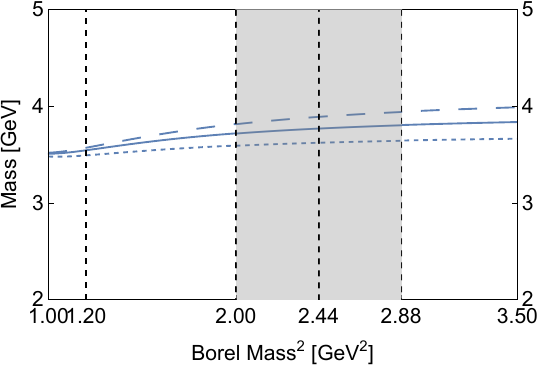}
\caption{
The extracted mass of the $J^{PC}=1^{--}$ $D\bar D$ state as a
function of the Borel parameter $M_B^2$ with
$s_0=17.5~{\rm GeV}^2$.
Although the OPE convergence criterion allows
$M_B^2\geq1.20~{\rm GeV}^2$, the extracted mass still shows a noticeable
dependence on $M_B^2$ in the low-$M_B^2$ region. A more stable behavior
is obtained for $M_B^2\geq2.0~{\rm GeV}^2$, which is therefore adopted
as the lower bound of the working region. The shaded region denotes the
final working window
$2.00~{\rm GeV}^2\leq M_B^2\leq2.88~{\rm GeV}^2$.
}
\label{fig:DDmass}
\end{figure}

We apply the same numerical procedure illustrated above to the remaining
$J^{P}=1^{-}$ $D^{(*)}\bar D^{(*)}$ channels constructed in
Sec.~\ref{sec:current}. The results for all the channels studied in this
work are summarized in Table~\ref{tab:results}. For the
$D\bar D^*/D^*\bar D$ system, the $J^{PC}=1^{-+}$ and
$J^{PC}=1^{--}$ channels have identical correlation functions, as
demonstrated in Sec.~\ref{sec:sumrule}, and consequently lead to the
same working regions and mass predictions in the present QCD sum rule
analysis.

In this numerical procedure, the Borel window for each channel is first
determined from the OPE convergence and pole dominance criteria. We then
examine the stability of the extracted mass with respect to the Borel
parameter and continuum threshold. For most channels, the extracted mass remains sufficiently stable
within these parameter regions, and no further restriction is needed. The only exception is the
$D\bar D$ channel. In this case, the OPE convergence criterion gives a
rather low lower bound, $M_B^2=1.20~{\rm GeV}^2$, while the extracted
mass still exhibits a noticeable dependence on $M_B^2$ in the region
below about $2.0~{\rm GeV}^2$. We therefore adopt a more conservative
lower bound,
\begin{equation}
M_B^2\geq2.0~{\rm GeV}^2,
\end{equation}
for this channel. The corresponding $M_B^2$ dependence of the extracted
mass is displayed in Fig.~\ref{fig:DDmass}.

\section{Summary and Discussion}
\label{sec:summary}

In this work, we have investigated the relative $P$-wave
$D^{(*)}\bar D^{(*)}$ meson-meson systems using the method of QCD
sum rules. We systematically constructed color-singlet meson-meson
interpolating currents in which a covariant derivative is explicitly
introduced between two ground-state charmed meson fields,
\begin{equation}
[\bar q\Gamma_1 c]\,
\overleftrightarrow D_\mu\,
[\bar c\Gamma_2 q] ,
\end{equation}
so that the derivative describes the relative orbital excitation between
the two meson clusters. Based on their Lorentz structures and charge
conjugation properties, we classified the possible $P$-wave
$D\bar D$, $D\bar D^*/D^*\bar D$, and $D^*\bar D^*$ configurations.
We then performed QCD sum rule analyses for the $J^P=1^-$ channels,
with the operator product expansion carried out up to dimension eight.

The central values obtained from the QCD sum rules lie close to the
corresponding charmed-meson thresholds and increase with the relevant
two-meson thresholds. The
$P$-wave $D\bar D$, $D\bar D^*/D^*\bar D$, and
$D^*\bar D^*$ configurations are obtained at approximately
\begin{eqnarray}
M_{D\bar D}&=&3.77^{+0.13}_{-0.16}~{\rm GeV},
\\
M_{D\bar D^*/D^*\bar D}&=&3.91^{+0.16}_{-0.19}~{\rm GeV},
\\
M_{D^*\bar D^*}&=&4.0\mbox{--}4.1~{\rm GeV}.
\end{eqnarray}
Taking into account the uncertainties inherent in the QCD sum rule
analysis, these predictions are compatible with configurations located
in the vicinity of the corresponding two-meson thresholds. In
particular, the obtained mass of the $P$-wave $D\bar D$ configuration
is compatible with the recently observed $X(3790)^-$ with
$J^P=1^-$, suggesting that this structure may be related to a relative
$P$-wave $D\bar D$ configuration. Similarly, the
$D\bar D^*/D^*\bar D$ result is compatible, within uncertainties, with
a structure located very close to the $D\bar D^*/D^*\bar D$ threshold,
as suggested by the $G(3900)$ structure. This is consistent with the proposal
that the $G(3900)$ may be interpreted as a $P$-wave
$D\bar D^*/D^*\bar D$ molecular resonance. These results motivate
further investigations of orbital excitations in charmed-meson systems.

An interesting feature appears in the
$D\bar D^*/D^*\bar D$ sector. Denoting the
$D\bar D^*$ and $D^*\bar D$ building-block currents by
$J_1$ and $J_2$, respectively, we find
\begin{equation}
\Pi_{11}(q^2)=\Pi_{22}(q^2),
\qquad
\Pi_{12}(q^2)=\Pi_{21}(q^2)=0,
\end{equation}
in our calculation. Consequently, the charge-conjugation eigenstates
with $J^{PC}=1^{-+}$ and $1^{--}$ lead to the same invariant
correlation function,
\begin{equation}
\Pi_{1^{-+}}(q^2)
=
\Pi_{1^{--}}(q^2).
\end{equation}
Therefore, these two channels lead to identical QCD sum rules and
degenerate mass predictions within the present framework. This
degeneracy applies only to the masses extracted from the present
two-point correlation functions. Since the two states have opposite
$C$ parities, their production mechanisms, decay properties, and
long-distance interactions may still be different and require separate
dynamical investigations.

The present study indicates that, besides the extensively investigated
$S$-wave hadronic molecules, relative $P$-wave charmed-meson systems
may constitute another interesting class of near-threshold hadronic
structures. Rather than implying a fixed excitation energy above the
corresponding thresholds, our results suggest the general pattern
\begin{equation}
M_{P\text{-wave}}
\sim
M_{\rm threshold},
\end{equation}
within the uncertainties of the present QCD sum rule analysis. The
central values typically lie within several tens of MeV of the
corresponding thresholds, while the uncertainties allow the physical
structures to appear very close to threshold. Further experimental
studies of the $X(3790)$, the $G(3900)$, and possible structures around
the $D^*\bar D^*$ thresholds will therefore be important for testing
this picture.

\section*{Acknowledgments}

This work is supported by
the National Natural Science Foundation of China under Grant No.~12505156 and 12075019,
the Jiangsu Provincial Double-Innovation Program under Grant No.~JSSCRC2021488, the Research Project
Supported by Shanxi Scholarship Council of China
and
the Fundamental Research Funds for the Central Universities.

\appendix
\begin{widetext}
\section{Spectral densities}
\label{app:ope}

In this appendix, we list the OPE expressions for the six
$J^P=1^-$ $D^{(*)}\bar D^{(*)}$ currents investigated in this work.
For compactness, we denote the corresponding OPE functions by
$\rho_i(s,M_B^2)$, with $i=1,\ldots,6$. Their correspondence to the
interpolating currents defined in Sec.~\ref{sec:current}, together with
the total spin $S$ of the two constituent mesons, is
\begin{eqnarray}
\rho_1 &\leftrightarrow& J^{D\bar D[1^{--}]}_{\alpha},
\qquad S=0 ,
\\
\rho_2 &\leftrightarrow& J^{D\bar D^*[1^{--}]}_{\alpha\beta},
\qquad S=1 ,
\\
\rho_3 &\leftrightarrow& J^{D\bar D^*[1^{-+}]}_{\alpha\beta},
\qquad S=1 ,
\\
\rho_4 &\leftrightarrow& J^{D^*\bar D^*[1^{--\prime}]}_{\alpha},
\qquad S=0 ,
\\
\rho_5 &\leftrightarrow& J^{D^*\bar D^*[1^{-+}]}_{\alpha},
\qquad S=1 ,
\\
\rho_6 &\leftrightarrow& J^{D^*\bar D^*[1^{--}]}_{\alpha},
\qquad S=2 .
\end{eqnarray}
In the following expressions, we define
\begin{equation}
\mathcal{F}(s)
=
(\alpha+\beta)m_c^2-\alpha\beta s,
\qquad
\mathcal{H}(s)
=
m_c^2-\alpha(1-\alpha)s ,
\end{equation}
with the integration limits
\begin{eqnarray}
\alpha_{\min}
&=&
\frac{1-\sqrt{1-4m_c^2/s}}{2},
\qquad
\alpha_{\max}
=
\frac{1+\sqrt{1-4m_c^2/s}}{2},
\\
\beta_{\min}
&=&
\frac{\alpha m_c^2}{\alpha s-m_c^2},
\qquad
\beta_{\max}
=
1-\alpha .
\end{eqnarray}

The OPE spectral density $\rho_{1}(s)$ extracted from the current $J^{D\bar D[1^{--}]}_{\alpha}$ is
\begin{equation}
	\rho_{1}(s,M_B^2) = \rho^{pert}_{1}(s) + \rho^{\qq}_{1}(s) + \rho^{\GGa}_{1}(s) + \rho^{\qGqa}_{1}(s) + \rho^{\qq^2}_{1}(s) + \rho^{\qq\qGqa_0}_{1}(s) + {\rho^{\qq\qGqa_2}_{1}(s) \over M_B^2} \, ,\label{ope:eta1}
\end{equation}
where
\begin{eqnarray}
	\nonumber \rho^{pert}_{1}(s) &=& \dab \Bigg\{ { \mathcal{F}(s)^5 } \times \frac{1}{81920 \alpha^4 \beta^4 \pi^6}  \times ( - 17 + 27 \alpha + 9 \alpha^2 - 19 \alpha^3 + 27 \beta - 30 \alpha \beta - 9 \alpha^2 \beta
	\non
	&+& 9 \beta^2 - 9 \alpha \beta^2 - 19 \beta^3) \Bigg\} \, ,
	\non
	\rho^{\qq}_{1}(s) &=& {\qq } \dab \Bigg\{ { \mathcal{F}(s)^3 }  \times \frac{m_c}{256 \alpha ^3 \beta ^3 \pi^4} \times ( - \alpha^2 + 10 \alpha \beta - 8 \alpha^2 \beta - \beta^2 - 8 \alpha \beta^2 ) \Bigg\} \, ,
	\non
	\rho^{\GGa}_{1}(s) &=& {\GGb }  \int^{\alpha_{max}}_{\alpha_{min}}d\alpha \int^{\beta_{max}}_{\beta_{min}}d\beta \Bigg\{ { \mathcal{F}(s)^3 } \times \frac{1}{32768 \alpha^3 \beta^3 \pi^6} \times ( - 5 \alpha + 6 \alpha^2 - \alpha^3 - 5 \beta -  20 \alpha \beta + 21 \alpha^2 \beta
	\non
	&+& 6 \beta^2 +  21 \alpha \beta^2 - \beta^3 ) + { \mathcal{F}(s)^3 } \times \frac{1}{8192 \alpha^3 \beta^3 \pi^2} \times ( \alpha^3 + \alpha^2 \beta + \alpha \beta^2 + \beta^3 ) + { \mathcal{F}(s)^2 } \times \frac{m_c^2}{98304 \alpha^4 \beta^4 \pi^6} 
	\non
	&\times&  ( - 17 \alpha^3 + 27 \alpha^4 + 9 \alpha^5 - 19 \alpha^6 + 27 \alpha^3 \beta - 30 \alpha^4 \beta - 9 \alpha^5 \beta + 9 \alpha^3 \beta^2 - 9 \alpha^4 \beta^2 - 17 \beta^3 + 27 \alpha \beta^3 + 9 \alpha^2 \beta^3
	\non
	&-& 38 \alpha^3 \beta^3 + 27 \beta^4 - 30 \alpha \beta^4 - 9 \alpha^2 \beta^4 + 9 \beta^5 - 9 \alpha \beta^5 - 19 \beta^6 ) \Bigg\} \, ,
	\non
	\rho^{\qGqa}_{1}(s) &=& {\qGqb } \int^{\alpha_{max}}_{\alpha_{min}}d\alpha  \Bigg\{ \int^{\beta_{max}}_{\beta_{min}}d\beta \Bigg[
	{ \mathcal{F}(s)^2 } \times \frac{m_c}{512 \alpha^3 \beta^3 \pi^4} \times ( - 3 \alpha^3 + 24 \alpha^2 \beta - 9 \alpha^3 \beta + 24 \alpha \beta^2 - 72 \alpha^2 \beta^2
    \non
    &-& 3 \beta^3 - 9 \alpha \beta^3 ) \Bigg] + { \mathcal{H}(s)^2 } \times \frac{m_c}{\pi^4} \times \frac{3 (1-2\alpha)^2}{1024 (1-\alpha)^2\alpha^2}  \Bigg\} \, ,
	\non
	\rho^{\qq^2}_{1}(s)&=& {\qq^2 } \int^{\alpha_{max}}_{\alpha_{min}}d\alpha \Bigg\{ { \mathcal{H}(s) } \times \frac{m_c^2}{\pi^2} \times \frac{(1-2\alpha)^2}{32 (1-\alpha)\alpha}  \Bigg\} \, ,
	\non
	\rho^{\qq\qGqa_0}_{1}(s)&=& {\qq\qGqb } \int^{\alpha_{max}}_{\alpha_{min}}d\alpha \Bigg\{ \frac{m_c^2}{\pi^2} \times \frac{ 1 - 8 \alpha + 20 \alpha^2 -24 \alpha^3 + 12 \alpha^4 }{64 (1-\alpha) \alpha} \Bigg\} \, ,
	\non
	\rho^{\qq\qGqa_2}_{1}(s)&=&{\qq\qGqb } \int^{1}_{0} \delta\left( s - {m_c^2 \over (1-\alpha) \alpha} \right) d\alpha  \Bigg\{ \frac{m_c^4}{\pi^2} \times \frac{(1-2\alpha)^2}{64 (-1 + \alpha) \alpha} \Bigg\} \, .
\end{eqnarray}

The OPE spectral density $\rho_{2}(s)$ extracted from the current $J^{D\bar D^*[1^{--}]}_{\alpha\beta}$ is
\begin{equation}
	\rho_{2}(s,M_B^2) = \rho^{pert}_{2}(s) + \rho^{\qq}_{2}(s) + \rho^{\GGa}_{2}(s) + \rho^{\qGqa}_{2}(s) + \rho^{\qq^2}_{2}(s) + \rho^{\qq\qGqa_0}_{2}(s) + {\rho^{\qq\qGqa_2}_{2}(s) \over M_B^2}\, ,
	\label{ope:eta5}
\end{equation}
where
\begin{eqnarray}
	\nonumber \rho^{pert}_{2}(s) &=& \dab \Bigg\{ { \mathcal{F}(s)^5 } \times \frac{1}{81920 \alpha^4 \beta^4 \pi^6}  \times ( - 5 + 21 \alpha^2 - 16 \alpha^3 + 6 \beta + 6 \alpha \beta - 18 \alpha^2 \beta + 9 \beta^2
	\non
	&-& 12 \alpha \beta^2 - 10 \beta^3) \Bigg\} \, ,
	\non
	\rho^{\qq}_{2}(s) &=& {\qq } \dab \Bigg\{ { \mathcal{F}(s)^3 }  \times \frac{m_c}{512 \alpha ^3 \beta ^3 \pi^4} \times ( - \alpha^3 + 9 \alpha \beta - 8 \alpha^2 \beta - \beta^2 - 7 \alpha \beta^2 ) \Bigg\} \, ,
	\non
	\rho^{\GGa}_{2}(s) &=& {\GGb }  \int^{\alpha_{max}}_{\alpha_{min}}d\alpha \int^{\beta_{max}}_{\beta_{min}}d\beta \Bigg\{ { \mathcal{F}(s)^3 } \times \frac{1}{98304 \alpha^3 \beta^3 \pi^6} \times ( - 3 \alpha - 6 \alpha^2 + 15  \alpha^3 + 5 \beta -  22 \alpha \beta + 15 \alpha^2 \beta
	\non
	&-& 8 \beta^2 +  33 \alpha \beta^2 + \beta^3 ) + { \mathcal{F}(s)^2 } \times \frac{m_c^2}{98304 \alpha^4 \beta^4 \pi^6} \times ( - 5 \alpha^3 + 21 \alpha^5 - 16 \alpha^6 + 6 \alpha^3 \beta + 6 \alpha^4 \beta - 18 \alpha^5 \beta 
    \non
	&+& 9 \alpha^3 \beta^2 - 12 \alpha^4 \beta^2 - 5 \beta^3 + 21 \alpha^2 \beta^3 - 26 \alpha^3 \beta^3 + 6 \beta^4 + 6 \alpha \beta^4 - 18 \alpha^2 \beta^4 + 9 \beta^5 - 12 \alpha \beta^5 - 10 \beta^6 ) \Bigg\} \, ,
	\non
	\rho^{\qGqa}_{2}(s) &=& {\qGqb } \int^{\alpha_{max}}_{\alpha_{min}}d\alpha  \Bigg\{ \int^{\beta_{max}}_{\beta_{min}}d\beta \Bigg[
	{ \mathcal{F}(s)^2 } \times \frac{m_c}{2048 \alpha^2 \beta^3 \pi^4} \times ( - 6 \alpha^3 + 36 \alpha \beta - 15 \alpha^2 \beta + 18 \beta^2 - 93 \alpha \beta^2
    \non
    &-& 6 \beta^3 ) \Bigg] + { \mathcal{H}(s)^2 } \times \frac{m_c}{\pi^4} \times \frac{3 (1-2\alpha)^2}{2048 (1-\alpha)^2\alpha^2}  \Bigg\} \, ,
	\non
	\rho^{\qq^2}_{2}(s)&=& {\qq^2 } \int^{\alpha_{max}}_{\alpha_{min}}d\alpha \Bigg\{ { \mathcal{H}(s) } \times \frac{m_c^2}{\pi^2} \times \frac{(1-2\alpha)^2}{64 (1-\alpha)\alpha}  \Bigg\} \, ,
	\non
	\rho^{\qq\qGqa_0}_{2}(s)&=& {\qq\qGqb } \int^{\alpha_{max}}_{\alpha_{min}}d\alpha \Bigg\{ \frac{m_c^2}{\pi^2} \times \frac{ 3 - 12 \alpha + 20 \alpha^2 - 12 \alpha^3 }{128 (-1+\alpha)} \Bigg\} \, ,
	\non
	\rho^{\qq\qGqa_2}_{2}(s)&=&{\qq\qGqb } \int^{1}_{0} \delta\left( s - {m_c^2 \over (1-\alpha) \alpha} \right) d\alpha  \Bigg\{ \frac{m_c^4}{\pi^2} \times \frac{(1-2\alpha)^2}{128 (-1 + \alpha) \alpha} \Bigg\} \, .
\end{eqnarray}

The OPE spectral density $\rho_{3}(s)$ extracted from the current $J^{D\bar D^*[1^{-+}]}_{\alpha\beta}$ is
\begin{equation}
	\rho_{3}(s,M_B^2) = \rho^{pert}_{3}(s) + \rho^{\qq}_{3}(s) + \rho^{\GGa}_{3}(s) + \rho^{\qGqa}_{3}(s) + \rho^{\qq^2}_{3}(s) + \rho^{\qq\qGqa_0}_{3}(s) + {\rho^{\qq\qGqa_2}_{3}(s) \over M_B^2}\, ,
	\label{ope:eta6}
\end{equation}
where
\begin{eqnarray}
	\nonumber \rho^{pert}_{3}(s) &=& \dab \Bigg\{ { \mathcal{F}(s)^5 } \times \frac{1}{81920 \alpha^4 \beta^4 \pi^6}  \times ( - 5 + 6 \alpha + 9 \alpha^2 - 10 \alpha^3 + 6 \alpha \beta - 12 \alpha^2 \beta + 21 \beta^2
	\non
	&-& 18 \alpha \beta^2 - 16 \beta^3) \Bigg\} \, ,
	\non
	\rho^{\qq}_{3}(s) &=& {\qq } \dab \Bigg\{ { \mathcal{F}(s)^3 }  \times \frac{m_c}{512 \alpha ^3 \beta ^3 \pi^4} \times ( - \alpha^2 + 9 \alpha \beta - 7 \alpha^2 \beta - 8 \alpha \beta^2 - \beta^3 ) \Bigg\} \, ,
	\non
	\rho^{\GGa}_{3}(s) &=& {\GGb }  \int^{\alpha_{max}}_{\alpha_{min}}d\alpha \int^{\beta_{max}}_{\beta_{min}}d\beta \Bigg\{ { \mathcal{F}(s)^3 } \times \frac{1}{98304 \alpha^3 \beta^3 \pi^6} \times ( 5 \alpha - 8 \alpha^2 + \alpha^3 - 3 \beta -  22 \alpha \beta + 33 \alpha^2 \beta
	\non
	&-& 6 \beta^2 + 15 \alpha \beta^2 + 15 \beta^3 ) + { \mathcal{F}(s)^2 } \times \frac{m_c^2}{98304 \alpha^4 \beta^4 \pi^6} \times ( - 5 \alpha^3 + 6 \alpha^4 + 9 \alpha^5 - 10 \alpha^6 + 6 \alpha^4 \beta - 12 \alpha^5 \beta 
    \non
    &+& 21 \alpha^3 \beta^2 - 18 \alpha^4 \beta^2 - 5 \beta^3 + 6 \alpha \beta^3 + 9 \alpha^2 \beta^3 - 26 \alpha^3 \beta^3 + 6 \alpha \beta^4 - 12 \alpha^2 \beta^4 + 21 \beta^5 - 18 \alpha \beta^5 - 16 \beta^6 ) \Bigg\} \, ,
	\non
	\rho^{\qGqa}_{3}(s) &=& {\qGqb } \int^{\alpha_{max}}_{\alpha_{min}}d\alpha  \Bigg\{ \int^{\beta_{max}}_{\beta_{min}}d\beta \Bigg[
	{ \mathcal{F}(s)^2 } \times \frac{m_c}{2048 \alpha^3 \beta^2 \pi^4} \times ( 18 \alpha^2 - 6 \alpha^3 + 36 \alpha \beta - 93 \alpha^2 \beta - 15 \alpha \beta^2 
    \non   
    &-& 6 \beta^3 ) + { \mathcal{H}(s)^2 } \times \frac{m_c}{\pi^4} \times \frac{3 (1-2\alpha)^2}{2048 (1-\alpha)^2\alpha^2}  \Bigg\} \, ,
	\non
	\rho^{\qq^2}_{3}(s)&=& {\qq^2 } \int^{\alpha_{max}}_{\alpha_{min}}d\alpha \Bigg\{ { \mathcal{H}(s) } \times \frac{m_c^2}{\pi^2} \times \frac{(1-2\alpha)^2}{64 (1-\alpha)\alpha}  \Bigg\} \, ,
	\non
	\rho^{\qq\qGqa_0}_{3}(s)&=& {\qq\qGqb } \int^{\alpha_{max}}_{\alpha_{min}}d\alpha \Bigg\{ \frac{m_c^2}{\pi^2} \times \frac{ 1 - 8 \alpha + 16 \alpha^2 - 12 \alpha^3 }{128 \alpha} \Bigg\} \, ,
	\non
	\rho^{\qq\qGqa_2}_{3}(s)&=&{\qq\qGqb } \int^{1}_{0} \delta\left( s - {m_c^2 \over (1-\alpha) \alpha} \right) d\alpha  \Bigg\{ \frac{m_c^4}{\pi^2} \times \frac{(1-2\alpha)^2}{128 (-1 + \alpha) \alpha} \Bigg\} \, .
\end{eqnarray}

The OPE spectral density $\rho_{4}(s)$ extracted from the current $J^{D^*\bar D^*[1^{--\prime}]}_{\alpha}$ is
\begin{equation}
	\rho_{4}(s,M_B^2) = \rho^{pert}_{4}(s) + \rho^{\qq}_{4}(s) + \rho^{\GGa}_{4}(s) + \rho^{\qGqa}_{4}(s) + \rho^{\qq^2}_{4}(s) + \rho^{\qq\qGqa_0}_{4}(s) + {\rho^{\qq\qGqa_2}_{4}(s) \over M_B^2}\, ,
	\label{ope:eta2}
\end{equation}
where
\begin{eqnarray}
	\nonumber \rho^{pert}_{4}(s) &=& \dab \Bigg\{ { \mathcal{F}(s)^5 } \times \frac{1}{40960 \alpha^4 \beta^4 \pi^6}  \times ( - 39 + 63 \alpha + 15 \alpha^2 - 39 \alpha^3 + 63 \beta - 66 \alpha \beta - 21 \alpha^2 \beta
	\non
	&+& 15 \beta^2 - 21 \alpha \beta^2 - 39 \beta^3) \Bigg\} \, ,
	\non
	\rho^{\qq}_{4}(s) &=& {\qq } \dab \Bigg\{ { \mathcal{F}(s)^3 }  \times \frac{m_c}{128 \alpha ^3 \beta ^3 \pi^4} \times ( - \alpha^2 + 10 \alpha \beta - 8 \alpha^2 \beta - \beta^2 - 8 \alpha \beta^2 ) \Bigg\} \, ,
	\non
	\rho^{\GGa}_{4}(s) &=& {\GGb }  \int^{\alpha_{max}}_{\alpha_{min}}d\alpha \int^{\beta_{max}}_{\beta_{min}}d\beta \Bigg\{ { \mathcal{F}(s)^3 } \times \frac{1}{49152 \alpha^3 \beta^3 \pi^6} \times ( - 3 \alpha + 2 \alpha^2 + \alpha^3 - 3 \beta + 4 \alpha \beta + 3 \alpha^2 \beta + 2 \beta^2
	\non
	&+& 3 \alpha \beta^2 + \beta^3 ) + { \mathcal{F}(s)^2 } \times \frac{m_c^2}{16384 \alpha^4 \beta^4 \pi^6} \times ( - 13 \alpha^3 + 21 \alpha^4 + 5 \alpha^5 - 13 \alpha^6 + 21 \alpha^3 \beta - 22 \alpha^4 \beta - 7 \alpha^5 \beta 
    \non
    &+& 5 \alpha^3 \beta^2 - 7 \alpha^4 \beta^2 - 13 \beta^3 + 21 \alpha \beta^3 + 5 \alpha^2 \beta^3 - 26 \alpha^3 \beta^3 + 21 \beta^4 - 22 \alpha \beta^4 - 7 \alpha^2 \beta^4 + 5 \beta^5 - 7 \alpha \beta^5 - 13 \beta^6 ) \Bigg\} \, ,
	\non
	\rho^{\qGqa}_{4}(s) &=& {\qGqb } \int^{\alpha_{max}}_{\alpha_{min}}d\alpha  \Bigg\{ \int^{\beta_{max}}_{\beta_{min}}d\beta \Bigg[
	{ \mathcal{F}(s)^2 } \times \frac{m_c}{256 \alpha^2 \beta^2 \pi^4} \times ( 9 \alpha - 3 \alpha^2 + 9 \beta - 36 \alpha \beta - 3 \beta^2 ) \Bigg] 
    \non
    &+& { \mathcal{H}(s)^2 } \times \frac{m_c}{\pi^4} \times \frac{3 (1-2\alpha)^2}{512 (1-\alpha)^2\alpha^2}  \Bigg\} \, ,
	\non
	\rho^{\qq^2}_{4}(s)&=& {\qq^2 } \int^{\alpha_{max}}_{\alpha_{min}}d\alpha \Bigg\{ { \mathcal{H}(s) } \times \frac{m_c^2}{\pi^2} \times \frac{(1-2\alpha)^2}{8 (1-\alpha)\alpha}  \Bigg\} \, ,
	\non
	\rho^{\qq\qGqa_0}_{4}(s)&=& {\qq\qGqb } \int^{\alpha_{max}}_{\alpha_{min}}d\alpha \Bigg\{ \frac{m_c^2}{\pi^2} \times \frac{ - 1 + 3 \alpha - 3 \alpha^2 }{4} \Bigg\} \, ,
	\non
	\rho^{\qq\qGqa_2}_{4}(s)&=&{\qq\qGqb } \int^{1}_{0} \delta\left( s - {m_c^2 \over (1-\alpha) \alpha} \right) d\alpha  \Bigg\{ \frac{m_c^4}{\pi^2} \times \frac{(1-2\alpha)^2}{16 (-1 + \alpha) \alpha} \Bigg\} \, .
\end{eqnarray}

The OPE spectral density $\rho_{5}(s)$ extracted from the current $J^{D^*\bar D^*[1^{-+}]}_{\alpha}$ is
\begin{equation}
	\rho_{5}(s,M_B^2) = \rho^{pert}_{5}(s) + \rho^{\qq}_{5}(s) + \rho^{\GGa}_{5}(s) + \rho^{\qGqa}_{5}(s) + \rho^{\qq^2}_{5}(s) + \rho^{\qq\qGqa_0}_{5}(s) + {\rho^{\qq\qGqa_2}_{5}(s) \over M_B^2} + {\rho^{\qq\qGqa_4}_{5}(s) \over M_B^4}\,,
	\label{ope:eta4}
\end{equation}
where
\begin{eqnarray}
    \nonumber \rho^{pert}_{5}(s) &=& \dab \Bigg\{ { \mathcal{F}(s)^5 } \times \frac{1}{163840 \alpha^4 \beta^4 \pi^6}  \times ( - 27 + 27 \alpha + 15 \alpha^2 - 39 \alpha^3 + 27 \beta + 78 \alpha \beta 
	\non
	&-& 69 \alpha^2 \beta + 15 \beta^2 - 69 \alpha \beta^2 - 39 \beta^3) \Bigg\} \, ,
	\non
	\rho^{\qq}_{5}(s) &=& {\qq } \dab \Bigg\{ { \mathcal{F}(s)^3 }  \times \frac{m_c}{512 \alpha^3 \beta^3 \pi^4} \times ( - \alpha^2 - \alpha^3 + 2 \alpha \beta + \alpha^2 \beta - \beta^2 + \alpha \beta^2 - \beta^3 ) \Bigg\} \, ,
	\non
	\rho^{\GGa}_{5}(s) &=& {\GGb }  \int^{\alpha_{max}}_{\alpha_{min}}d\alpha \Bigg\{ \int^{\beta_{max}}_{\beta_{min}}d\beta \Bigg[ { \mathcal{F}(s)^3 } \times \frac{1}{196608 \alpha^3 \beta^3 \pi^6} \times ( 9 \alpha - 46 \alpha^2 + 9 \alpha^3 + 9 \beta - 28 \alpha \beta 
    \non
    &+& 75 \alpha^2 \beta - 46 \beta^2 + 75 \alpha \beta^2 + 9 \beta^3 ) + { \mathcal{F}(s)^2 } \times \frac{m_c^2}{65536 \alpha^4 \beta^4 \pi^6} \times ( - 9 \alpha^3 + 9 \alpha^4 + 5 \alpha^5 - 13 \alpha^6 + 9 \alpha^3 \beta 
    \non
    &+& 26 \alpha^4 \beta - 23 \alpha^5 \beta + 5 \alpha^3 \beta^2 - 23 \alpha^4 \beta^2 - 9 \beta^3 + 9 \alpha \beta^3 + 5 \alpha^2 \beta^3 - 26 \alpha^3 \beta^3 + 9 \beta^4 + 26 \alpha \beta^4 - 23 \alpha^2 \beta^4
    \non
    &+& 5 \beta^5 - 23 \alpha \beta^5 - 13 \beta^6 ) \Bigg] + { \mathcal{H}(s)^3 } \times \frac{1}{\pi^6} \times \frac{(1-2\alpha)^2}{24576 (-1+\alpha)^3\alpha^3}  \Bigg\} \, ,
	\non
	\rho^{\qGqa}_{5}(s) &=& {\qGqb } \int^{\alpha_{max}}_{\alpha_{min}}d\alpha  \Bigg\{ \int^{\beta_{max}}_{\beta_{min}}d\beta \Bigg[
	{ \mathcal{F}(s)^2 } \times \frac{m_c}{2048 \alpha^2 \beta^2 \pi^4} \times ( 6 \alpha + 3 \alpha^2 + 6 \beta + 18 \alpha \beta + 3 \beta^2 ) \Bigg] 
    \non
    &+& { \mathcal{H}(s)^2 } \times \frac{m_c}{\pi^4} \times \frac{3 (1-2\alpha)^2}{1024 (1-\alpha)^2\alpha^2}  \Bigg\} \, ,
	\non
	\rho^{\qq^2}_{5}(s)&=& {\qq^2 } \int^{\alpha_{max}}_{\alpha_{min}}d\alpha \Bigg\{ { \mathcal{H}(s) } \times \frac{m_c^2}{\pi^2} \times \frac{5 (1-2\alpha)^2}{64 (1-\alpha)\alpha} + \frac{m_c^4}{\pi^2} \times \frac{(1-2\alpha)^2}{32 (-1+\alpha)\alpha} \Bigg\} \, ,
	\non
	\rho^{\qq\qGqa_0}_{5}(s)&=& {\qq\qGqb } \int^{\alpha_{max}}_{\alpha_{min}}d\alpha \Bigg\{ \frac{m_c^2}{\pi^2} \times \frac{ 3(-1 + 3 \alpha - 3 \alpha^2) }{32} \Bigg\} \, ,
	\non
	\rho^{\qq\qGqa_2}_{5}(s)&=&{\qq\qGqb } \int^{1}_{0} \delta\left( s - {m_c^2 \over (1-\alpha) \alpha} \right) d\alpha  \Bigg\{ \frac{m_c^4}{\pi^2} \times \frac{7(1-2\alpha)^2}{128 (-1 + \alpha) \alpha} \Bigg\} \, ,
    \non
	\rho^{\qq\qGqa_4}_{5}(s)&=&{\qq\qGqb } \int^{1}_{0} \delta\left( s - {m_c^2 \over (1-\alpha) \alpha} \right) d\alpha  \Bigg\{ \frac{m_c^6}{\pi^2} \times - \frac{(1-2\alpha)^2}{64 (1 - \alpha)^2 \alpha^2} \Bigg\} \, .
\end{eqnarray}

The OPE spectral density $\rho_{6}(s)$ extracted from the current $J^{D^*\bar D^*[1^{--}]}_{\alpha}$ is
\begin{equation}
	\rho_{6}(s,M_B^2) = \rho^{pert}_{6}(s) + \rho^{\qq}_{6}(s) + \rho^{\GGa}_{6}(s) + \rho^{\qGqa}_{6}(s) + \rho^{\qq^2}_{6}(s) + \rho^{\qq\qGqa_0}_{6}(s) + {\rho^{\qq\qGqa_2}_{6}(s) \over M_B^2} + {\rho^{\qq\qGqa_4}_{6}(s) \over M_B^4}\,,
	\label{ope:eta3}
\end{equation}
where
\begin{eqnarray}
    \nonumber \rho^{pert}_{6}(s) &=& \dab \Bigg\{ { \mathcal{F}(s)^5 } \times \frac{1}{655360 \alpha^4 \beta^4 \pi^6}  \times ( - 207 + 351 \alpha - 57 \alpha^2 - 183 \alpha^3 + 351 \beta - 210 \alpha \beta 
	\non
	&-& 69 \alpha^2 \beta - 57 \beta^2 - 69 \alpha \beta^2 - 183 \beta^3) \Bigg\} \, ,
	\non
	\rho^{\qq}_{6}(s) &=& {\qq } \dab \Bigg\{ { \mathcal{F}(s)^3 }  \times \frac{m_c}{2048 \alpha ^3 \beta ^3 \pi^4} \times ( - \alpha^2 - 8 \alpha^3 + 10 \alpha \beta - \beta^2 - 8 \beta^3 ) \Bigg\} \, ,
	\non
	\rho^{\GGa}_{6}(s) &=& {\GGb }  \int^{\alpha_{max}}_{\alpha_{min}}d\alpha \Bigg\{ \int^{\beta_{max}}_{\beta_{min}}d\beta \Bigg[ { \mathcal{F}(s)^3 } \times \frac{1}{786432 \alpha^3 \beta^3 \pi^6} \times ( - 27 \alpha - 30 \alpha^2 - 39 \alpha^3 - 27 \beta + 68 \alpha \beta 
    \non
    &+& 139 \alpha^2 \beta - 30 \beta^2 + 139 \alpha \beta^2 - 39 \beta^3 ) + { \mathcal{F}(s)^2 } \times \frac{m_c^2}{262144 \alpha^4 \beta^4 \pi^6} \times ( - 69 \alpha^3 + 117 \alpha^4 - 19 \alpha^5 - 61 \alpha^6 
    \non
    &+& 117 \alpha^3 \beta - 70 \alpha^4 \beta - 23 \alpha^5 \beta - 19 \alpha^3 \beta^2 - 23 \alpha^4 \beta^2 - 69 \beta^3 + 117 \alpha \beta^3 - 19 \alpha^2 \beta^3 - 122 \alpha^3 \beta^3 + 117 \beta^4 
    \non
    &-& 70 \alpha \beta^4 - 23 \alpha^2 \beta^4 - 19 \beta^5 - 23 \alpha \beta^5 - 61 \beta^6 ) \Bigg] + { \mathcal{H}(s)^3 } \times \frac{1}{\pi^6} \times \frac{(1-2\alpha)^2}{24576 (-1+\alpha)^3\alpha^3}  \Bigg\} \, ,
	\non
	\rho^{\qGqa}_{6}(s) &=& {\qGqb } \int^{\alpha_{max}}_{\alpha_{min}}d\alpha  \Bigg\{ \int^{\beta_{max}}_{\beta_{min}}d\beta \Bigg[
	{ \mathcal{F}(s)^2 } \times \frac{m_c}{4096 \alpha^2 \beta^2 \pi^4} \times ( 15 \alpha - 3 \alpha^2 + 15 \beta + 84 \alpha \beta - 3 \beta^2 ) \Bigg] 
    \non
    &+& { \mathcal{H}(s)^2 } \times \frac{m_c}{\pi^4} \times \frac{27 (1-2\alpha)^2}{8192 (1-\alpha)^2\alpha^2}  \Bigg\} \, ,
	\non
	\rho^{\qq^2}_{6}(s)&=& {\qq^2 } \int^{\alpha_{max}}_{\alpha_{min}}d\alpha \Bigg\{ { \mathcal{H}(s) } \times \frac{m_c^2}{\pi^2} \times \frac{13 (1-2\alpha)^2}{128 (1-\alpha)\alpha} + \frac{m_c^4}{\pi^2} \times \frac{(1-2\alpha)^2}{32 (-1+\alpha)\alpha} \Bigg\} \, ,
	\non
	\rho^{\qq\qGqa_0}_{6}(s)&=& {\qq\qGqb } \int^{\alpha_{max}}_{\alpha_{min}}d\alpha \Bigg\{ \frac{m_c^2}{\pi^2} \times \frac{ 9(-1 + 3 \alpha - 3 \alpha^2) }{64} \Bigg\} \, ,
	\non
	\rho^{\qq\qGqa_2}_{6}(s)&=&{\qq\qGqb } \int^{1}_{0} \delta\left( s - {m_c^2 \over (1-\alpha) \alpha} \right) d\alpha  \Bigg\{ \frac{m_c^4}{\pi^2} \times \frac{17(1-2\alpha)^2}{256 (-1 + \alpha) \alpha} \Bigg\} \, ,
    \non
	\rho^{\qq\qGqa_4}_{6}(s)&=&{\qq\qGqb } \int^{1}_{0} \delta\left( s - {m_c^2 \over (1-\alpha) \alpha} \right) d\alpha  \Bigg\{ \frac{m_c^6}{\pi^2} \times - \frac{(1-2\alpha)^2}{64 (1 - \alpha)^2 \alpha^2} \Bigg\} \, .
\end{eqnarray}

\end{widetext}

\bibliographystyle{elsarticle-num}
\bibliography{ref}

\begin{thebibliography}{10}
\expandafter\ifx\csname url\endcsname\relax
  \def\url#1{\texttt{#1}}\fi
\expandafter\ifx\csname urlprefix\endcsname\relax\def\urlprefix{URL }\fi
\expandafter\ifx\csname href\endcsname\relax
  \def\href#1#2{#2} \def\path#1{#1}\fi

\bibitem{pdg}
S.~Navas, et~al., {Review of particle physics}, Phys. Rev. D 110~(3) (2024) 030001.
\newblock \href {https://doi.org/10.1103/PhysRevD.110.030001} {\path{doi:10.1103/PhysRevD.110.030001}}.

\bibitem{Belle:2003nnu}
S.~K. Choi, et~al., {Observation of a Narrow Charmoniumlike State in Exclusive $B^\pm \to K^\pm \pi^+ \pi^- J/\psi$ Decays}, Phys. Rev. Lett. 91 (2003) 262001.
\newblock \href {http://arxiv.org/abs/hep-ex/0309032} {\path{arXiv:hep-ex/0309032}}, \href {https://doi.org/10.1103/PhysRevLett.91.262001} {\path{doi:10.1103/PhysRevLett.91.262001}}.

\bibitem{Chen:2016qju}
H.-X. Chen, W.~Chen, X.~Liu, S.-L. Zhu, {The hidden-charm pentaquark and tetraquark states}, Phys. Rept. 639 (2016) 1--121.
\newblock \href {http://arxiv.org/abs/1601.02092} {\path{arXiv:1601.02092}}, \href {https://doi.org/10.1016/j.physrep.2016.05.004} {\path{doi:10.1016/j.physrep.2016.05.004}}.

\bibitem{Liu:2019zoy}
Y.-R. Liu, H.-X. Chen, W.~Chen, X.~Liu, S.-L. Zhu, {Pentaquark and Tetraquark States}, Prog. Part. Nucl. Phys. 107 (2019) 237--320.
\newblock \href {http://arxiv.org/abs/1903.11976} {\path{arXiv:1903.11976}}, \href {https://doi.org/10.1016/j.ppnp.2019.04.003} {\path{doi:10.1016/j.ppnp.2019.04.003}}.

\bibitem{Chen:2022asf}
H.-X. Chen, W.~Chen, X.~Liu, Y.-R. Liu, S.-L. Zhu, {An updated review of the new hadron states}, Rept. Prog. Phys. 86~(2) (2023) 026201.
\newblock \href {http://arxiv.org/abs/2204.02649} {\path{arXiv:2204.02649}}, \href {https://doi.org/10.1088/1361-6633/aca3b6} {\path{doi:10.1088/1361-6633/aca3b6}}.

\bibitem{Luo:2025sns}
X.~Luo, S.-W. Zhang, H.-X. Chen, A.~Hosaka, N.~Su, H.-M. Yang, {A short review on QCD sum rule studies of $P$-wave single heavy baryons}, AAPPS Bull. 36~(1) (2026) 9.
\newblock \href {http://arxiv.org/abs/2510.13013} {\path{arXiv:2510.13013}}, \href {https://doi.org/10.1007/s43673-026-00188-8} {\path{doi:10.1007/s43673-026-00188-8}}.

\bibitem{Hosaka:2016pey}
A.~Hosaka, T.~Iijima, K.~Miyabayashi, Y.~Sakai, S.~Yasui, {Exotic hadrons with heavy flavors: X, Y, Z, and related states}, PTEP 2016~(6) (2016) 062C01.
\newblock \href {http://arxiv.org/abs/1603.09229} {\path{arXiv:1603.09229}}, \href {https://doi.org/10.1093/ptep/ptw045} {\path{doi:10.1093/ptep/ptw045}}.

\bibitem{Richard:2016eis}
J.-M. Richard, {Exotic Hadrons: Review and Perspectives}, Few Body Syst. 57~(12) (2016) 1185--1212.
\newblock \href {http://arxiv.org/abs/1606.08593} {\path{arXiv:1606.08593}}, \href {https://doi.org/10.1007/s00601-016-1159-0} {\path{doi:10.1007/s00601-016-1159-0}}.

\bibitem{Lebed:2016hpi}
R.~F. Lebed, R.~E. Mitchell, E.~S. Swanson, {Heavy-quark QCD exotica}, Prog. Part. Nucl. Phys. 93 (2017) 143--194.
\newblock \href {http://arxiv.org/abs/1610.04528} {\path{arXiv:1610.04528}}, \href {https://doi.org/10.1016/j.ppnp.2016.11.003} {\path{doi:10.1016/j.ppnp.2016.11.003}}.

\bibitem{Esposito:2016noz}
A.~Esposito, A.~Pilloni, A.~D. Polosa, {Multiquark resonances}, Phys. Rept. 668 (2017) 1--97.
\newblock \href {http://arxiv.org/abs/1611.07920} {\path{arXiv:1611.07920}}, \href {https://doi.org/10.1016/j.physrep.2016.11.002} {\path{doi:10.1016/j.physrep.2016.11.002}}.

\bibitem{Ali:2017jda}
A.~Ali, J.~S. Lange, S.~Stone, {Exotics: Heavy pentaquarks and tetraquarks}, Prog. Part. Nucl. Phys. 97 (2017) 123--198.
\newblock \href {http://arxiv.org/abs/1706.00610} {\path{arXiv:1706.00610}}, \href {https://doi.org/10.1016/j.ppnp.2017.08.003} {\path{doi:10.1016/j.ppnp.2017.08.003}}.

\bibitem{Guo:2017jvc}
F.-K. Guo, C.~Hanhart, U.-G. Mei\ss{}ner, Q.~Wang, Q.~Zhao, B.-S. Zou, {Hadronic molecules}, Rev. Mod. Phys. 90~(1) (2018) 015004.
\newblock \href {http://arxiv.org/abs/1705.00141} {\path{arXiv:1705.00141}}, \href {https://doi.org/10.1103/RevModPhys.90.015004} {\path{doi:10.1103/RevModPhys.90.015004}}.

\bibitem{Olsen:2017bmm}
S.~L. Olsen, T.~Skwarnicki, D.~Zieminska, {Nonstandard heavy mesons and baryons: Experimental evidence}, Rev. Mod. Phys. 90~(1) (2018) 015003.
\newblock \href {http://arxiv.org/abs/1708.04012} {\path{arXiv:1708.04012}}, \href {https://doi.org/10.1103/RevModPhys.90.015003} {\path{doi:10.1103/RevModPhys.90.015003}}.

\bibitem{Karliner:2017qhf}
M.~Karliner, J.~L. Rosner, T.~Skwarnicki, {Multiquark States}, Ann. Rev. Nucl. Part. Sci. 68 (2018) 17--44.
\newblock \href {http://arxiv.org/abs/1711.10626} {\path{arXiv:1711.10626}}, \href {https://doi.org/10.1146/annurev-nucl-101917-020902} {\path{doi:10.1146/annurev-nucl-101917-020902}}.

\bibitem{Guo:2019twa}
F.-K. Guo, X.-H. Liu, S.~Sakai, {Threshold cusps and triangle singularities in hadronic reactions}, Prog. Part. Nucl. Phys. 112 (2020) 103757.
\newblock \href {http://arxiv.org/abs/1912.07030} {\path{arXiv:1912.07030}}, \href {https://doi.org/10.1016/j.ppnp.2020.103757} {\path{doi:10.1016/j.ppnp.2020.103757}}.

\bibitem{Brambilla:2019esw}
N.~Brambilla, S.~Eidelman, C.~Hanhart, A.~Nefediev, C.-P. Shen, C.~E. Thomas, A.~Vairo, C.-Z. Yuan, {The $XYZ$ states: Experimental and theoretical status and perspectives}, Phys. Rept. 873 (2020) 1--154.
\newblock \href {http://arxiv.org/abs/1907.07583} {\path{arXiv:1907.07583}}, \href {https://doi.org/10.1016/j.physrep.2020.05.001} {\path{doi:10.1016/j.physrep.2020.05.001}}.

\bibitem{Yang:2020atz}
G.~Yang, J.~Ping, J.~Segovia, {Tetra- and Penta-Quark Structures in the Constituent Quark Model}, Symmetry 12~(11) (2020) 1869.
\newblock \href {http://arxiv.org/abs/2009.00238} {\path{arXiv:2009.00238}}, \href {https://doi.org/10.3390/sym12111869} {\path{doi:10.3390/sym12111869}}.

\bibitem{Fang:2021wes}
S.-S. Fang, B.~Kubis, A.~Kup{\'s}{\'c}, {What can we learn about light-meson interactions at electron-positron colliders?}, Prog. Part. Nucl. Phys. 120 (2021) 103884.
\newblock \href {http://arxiv.org/abs/2102.05922} {\path{arXiv:2102.05922}}, \href {https://doi.org/10.1016/j.ppnp.2021.103884} {\path{doi:10.1016/j.ppnp.2021.103884}}.

\bibitem{Jin:2021vct}
S.~Jin, X.~Shen, {Highlights of light meson spectroscopy at the BESIII experiment}, Natl. Sci. Rev. 8~(11) (2021) nwab198.
\newblock \href {https://doi.org/10.1093/nsr/nwab198} {\path{doi:10.1093/nsr/nwab198}}.

\bibitem{Meng:2022ozq}
L.~Meng, B.~Wang, G.-J. Wang, S.-L. Zhu, {Chiral perturbation theory for heavy hadrons and chiral effective field theory for heavy hadronic molecules}, Phys. Rept. 1019 (2023) 1--149.
\newblock \href {http://arxiv.org/abs/2204.08716} {\path{arXiv:2204.08716}}, \href {https://doi.org/10.1016/j.physrep.2023.04.003} {\path{doi:10.1016/j.physrep.2023.04.003}}.

\bibitem{Liu:2024uxn}
M.-Z. Liu, Y.-W. Pan, Z.-W. Liu, T.-W. Wu, J.-X. Lu, L.-S. Geng, {Three ways to decipher the nature of exotic hadrons: Multiplets, three-body hadronic molecules, and correlation functions}, Phys. Rept. 1108 (2025) 1--108.
\newblock \href {http://arxiv.org/abs/2404.06399} {\path{arXiv:2404.06399}}, \href {https://doi.org/10.1016/j.physrep.2024.12.001} {\path{doi:10.1016/j.physrep.2024.12.001}}.

\bibitem{Wang:2025dur}
X.~Wang, X.~Liu, Y.~Gao, {Colloquium: Hadron production in open-charm meson pairs at $e^+e^-$ colliders}, Rev. Mod. Phys. 98~(2) (2026) 021001.
\newblock \href {http://arxiv.org/abs/2502.15117} {\path{arXiv:2502.15117}}, \href {https://doi.org/10.1103/2mrp-chly} {\path{doi:10.1103/2mrp-chly}}.

\bibitem{Dai:2026fkg}
X.~Dai, S.~Jia, A.~Nefediev, J.~Nieves, C.~Shen, L.~Zhang, {Exotic hadrons associated with $b$-quark}, Phys. Rept. 1191 (2026) 1--62.
\newblock \href {http://arxiv.org/abs/2603.09315} {\path{arXiv:2603.09315}}, \href {https://doi.org/10.1016/j.physrep.2026.06.002} {\path{doi:10.1016/j.physrep.2026.06.002}}.

\bibitem{Bai:2026atm}
Z.-Y. Bai, D.-Y. Chen, Q.~Huang, X.~Liu, S.-Q. Luo, J.-Z. Wang, {Unquenched charmonium and beyond}, Phys. Rept. 1204 (2027) 1--162.
\newblock \href {http://arxiv.org/abs/2602.19887} {\path{arXiv:2602.19887}}, \href {https://doi.org/10.1016/j.physrep.2026.08.002} {\path{doi:10.1016/j.physrep.2026.08.002}}.

\bibitem{Du:2016qcr}
M.-L. Du, U.-G. Mei{\ss}ner, Q.~Wang, {$P$-wave coupled channel effects in electron-positron annihilation}, Phys. Rev. D 94~(9) (2016) 096006.
\newblock \href {http://arxiv.org/abs/1608.02537} {\path{arXiv:1608.02537}}, \href {https://doi.org/10.1103/PhysRevD.94.096006} {\path{doi:10.1103/PhysRevD.94.096006}}.

\bibitem{BESIII:2024ths}
M.~Ablikim, et~al., {Precise Measurement of Born Cross Sections for e+e-{\textrightarrow}DD{\textasciimacron} at s=3.80-4.95{\,}{\,}GeV}, Phys. Rev. Lett. 133~(8) (2024) 081901.
\newblock \href {http://arxiv.org/abs/2402.03829} {\path{arXiv:2402.03829}}, \href {https://doi.org/10.1103/PhysRevLett.133.081901} {\path{doi:10.1103/PhysRevLett.133.081901}}.

\bibitem{Lin:2024qcq}
Z.-Y. Lin, J.-Z. Wang, J.-B. Cheng, L.~Meng, S.-L. Zhu, {Identification of the $G(3900)$ as the P-wave $D\bar{D}^*/\bar{D}D^*$ resonance}, Phys. Rev. Lett. 133~(24) (2024) 241903.
\newblock \href {http://arxiv.org/abs/2403.01727} {\path{arXiv:2403.01727}}, \href {https://doi.org/10.1103/PhysRevLett.133.241903} {\path{doi:10.1103/PhysRevLett.133.241903}}.

\bibitem{Huang:2025rvj}
Y.~Huang, X.~Chen, {Direct Evidence for the $\bar{D}D^*/D\bar{D}^*$ Molecular Nature of $G(3900)$ Through Triangular Singularity Mechanisms} (1 2025).
\newblock \href {http://arxiv.org/abs/2501.10992} {\path{arXiv:2501.10992}}.

\bibitem{Cao:2025okk}
J.~Cao, W.-C. Zhang, Z.-L. She, A.-K. Lei, J.-P. Zhang, H.~Zheng, D.-M. Zhou, Y.-L. Yan, Z.-Q. Wang, B.-H. Sa, {Charmonium-like exotic hadron productions in e+e- collisions at the BESIII energy with the PACIAE model}, Phys. Rev. D 112~(1) (2025) 014033.
\newblock \href {http://arxiv.org/abs/2502.16822} {\path{arXiv:2502.16822}}, \href {https://doi.org/10.1103/1gtz-xkgh} {\path{doi:10.1103/1gtz-xkgh}}.

\bibitem{Lu:2025zae}
J.-L. Lu, M.~Song, P.~Wang, J.-Y. Guo, G.~Li, X.~Luo, {The bound and resonant states of $D^{(*)}D^{(*)}$ and $D^{(*)}{\bar{D}}^{(*)}$ with the complex scaling method}, Eur. Phys. J. C 85~(8) (2025) 920.
\newblock \href {http://arxiv.org/abs/2503.05131} {\path{arXiv:2503.05131}}, \href {https://doi.org/10.1140/epjc/s10052-025-14649-1} {\path{doi:10.1140/epjc/s10052-025-14649-1}}.

\bibitem{Chen:2025gxe}
X.-X. Chen, Z.-M. Ding, J.~He, {Pole trajectories from S- and P-wave DD{\textasciimacron}* interactions}, Phys. Rev. D 111~(11) (2025) 114008.
\newblock \href {http://arxiv.org/abs/2504.15534} {\path{arXiv:2504.15534}}, \href {https://doi.org/10.1103/36vy-r2z8} {\path{doi:10.1103/36vy-r2z8}}.

\bibitem{Ye:2025ywy}
Q.~Ye, Z.~Zhang, M.-L. Du, U.-G. Mei{\ss}ner, P.-Y. Niu, Q.~Wang, {Resonance parameters of the vector charmoniumlike state G(3900)}, Phys. Rev. D 112~(1) (2025) 016015.
\newblock \href {http://arxiv.org/abs/2504.17431} {\path{arXiv:2504.17431}}, \href {https://doi.org/10.1103/qq61-ncln} {\path{doi:10.1103/qq61-ncln}}.

\bibitem{Liu:2025sjz}
S.-D. Liu, Q.~Wu, G.~Li, {Hadronic decays of possible pseudoscalar P-wave DD{\textasciimacron}*/D{\textasciimacron}D* molecular state}, Phys. Rev. D 112~(7) (2025) 074002.
\newblock \href {http://arxiv.org/abs/2506.18273} {\path{arXiv:2506.18273}}, \href {https://doi.org/10.1103/t3q7-fml8} {\path{doi:10.1103/t3q7-fml8}}.

\bibitem{Chen:2025mgg}
K.~Chen, J.-Z. Wang, {$D$-wave dimeson resonance interpretation to the $Z_c/Z_{cs}/Z_b$ states} (8 2025).
\newblock \href {http://arxiv.org/abs/2508.11127} {\path{arXiv:2508.11127}}.

\bibitem{Qian:2025zyp}
R.-Q. Qian, X.~Liu, {Unified coupled-channel description for the five near-threshold structures {\ensuremath{\psi}}(3770), G(3900), R(3760), R(3780) and R(3810) from e+e- annihilation}, Phys. Rev. D 112~(9) (2025) L091502.
\newblock \href {http://arxiv.org/abs/2509.17679} {\path{arXiv:2509.17679}}, \href {https://doi.org/10.1103/c5vs-2l1k} {\path{doi:10.1103/c5vs-2l1k}}.

\bibitem{Wang:2026mzg}
J.~Wang, Q.~Zhao, {Study of the e+e-{\textrightarrow}J/{\ensuremath{\psi}}{\ensuremath{\pi}}+{\ensuremath{\pi}}- lineshape near the D*D{\textasciimacron}+c.c. threshold and possible signals for exotic hidden charm states}, Phys. Rev. D 114~(5) (2026) 054020.
\newblock \href {http://arxiv.org/abs/2603.19550} {\path{arXiv:2603.19550}}, \href {https://doi.org/10.1103/zptx-65dy} {\path{doi:10.1103/zptx-65dy}}.

\bibitem{Lu:2026zcs}
Q.~Lu, C.~Cheng, Y.~Huang, {Searching for the $G(3900)$ via the $K^- p \to D_s^- \Lambda_c^+ G(3900)^0$ reaction} (6 2026).
\newblock \href {http://arxiv.org/abs/2606.29359} {\path{arXiv:2606.29359}}.

\bibitem{Abreu:2026ngb}
L.~M. Abreu, P.~Brand{\~a}o, R.~O. Magalh{\~a}es, F.~S. Navarra, {Probing the $G(3900)$ State in Heavy-Ion Collisions} (6 2026).
\newblock \href {http://arxiv.org/abs/2606.13909} {\path{arXiv:2606.13909}}.

\bibitem{BESIII:2026emz}
M.~Ablikim, et~al., {Observation of a Charged Charmoniumlike Structure in $e^+e^-\rightarrow D^0D^-\pi^++c.c.$} (7 2026).
\newblock \href {http://arxiv.org/abs/2607.29138} {\path{arXiv:2607.29138}}.

\bibitem{Cui:2013yva}
C.-Y. Cui, Y.-L. Liu, W.-B. Chen, M.-Q. Huang, {Could $Z_{c}(3900)$ be a $I^{G}J^{P}=1^{+}1^{+}$ $D^{*}\bar{D}$ molecular state?}, J. Phys. G 41 (2014) 075003.
\newblock \href {http://arxiv.org/abs/1304.1850} {\path{arXiv:1304.1850}}, \href {https://doi.org/10.1088/0954-3899/41/7/075003} {\path{doi:10.1088/0954-3899/41/7/075003}}.

\bibitem{Chen:2013omd}
W.~Chen, T.~G. Steele, M.-L. Du, S.-L. Zhu, {$D^*\bar D^*$ molecule interpretation of $Z_c(4025)$}, Eur. Phys. J. C 74~(2) (2014) 2773.
\newblock \href {http://arxiv.org/abs/1308.5060} {\path{arXiv:1308.5060}}, \href {https://doi.org/10.1140/epjc/s10052-014-2773-y} {\path{doi:10.1140/epjc/s10052-014-2773-y}}.

\bibitem{Wang:2013daa}
Z.-G. Wang, T.~Huang, {Possible assignments of the $X(3872)$, $Z_c(3900)$ and $Z_b(10610)$ as axial-vector molecular states}, Eur. Phys. J. C 74~(5) (2014) 2891.
\newblock \href {http://arxiv.org/abs/1312.7489} {\path{arXiv:1312.7489}}, \href {https://doi.org/10.1140/epjc/s10052-014-2891-6} {\path{doi:10.1140/epjc/s10052-014-2891-6}}.

\bibitem{Chen:2015kpa}
H.-X. Chen, W.~Chen, Q.~Mao, A.~Hosaka, X.~Liu, S.-L. Zhu, {$P$-wave charmed baryons from QCD sum rules}, Phys. Rev. D 91~(5) (2015) 054034.
\newblock \href {http://arxiv.org/abs/1502.01103} {\path{arXiv:1502.01103}}, \href {https://doi.org/10.1103/PhysRevD.91.054034} {\path{doi:10.1103/PhysRevD.91.054034}}.

\bibitem{Su:2022eun}
N.~Su, H.-X. Chen, {$S$- and $P$-wave fully strange tetraquark states from QCD sum rules}, Phys. Rev. D 106~(1) (2022) 014023.
\newblock \href {http://arxiv.org/abs/2204.13959} {\path{arXiv:2204.13959}}, \href {https://doi.org/10.1103/PhysRevD.106.014023} {\path{doi:10.1103/PhysRevD.106.014023}}.

\bibitem{Wang:2018ejf}
Z.-G. Wang, {Analysis of the vector tetraquark states with P-waves between the diquarks and antidiquarks via the QCD sum rules}, Eur. Phys. J. C 79~(1) (2019) 29.
\newblock \href {http://arxiv.org/abs/1811.02726} {\path{arXiv:1811.02726}}, \href {https://doi.org/10.1140/epjc/s10052-019-6568-z} {\path{doi:10.1140/epjc/s10052-019-6568-z}}.

\bibitem{Xin:2022qnv}
Q.~Xin, Z.-G. Wang, {Fully-light vector tetraquark states with explicit P-wave via QCD sum rules*}, Chin. Phys. C 48~(3) (2024) 033104.
\newblock \href {http://arxiv.org/abs/2211.14993} {\path{arXiv:2211.14993}}, \href {https://doi.org/10.1088/1674-1137/ad181c} {\path{doi:10.1088/1674-1137/ad181c}}.

\bibitem{Xi:2024cvy}
H.-Z. Xi, H.-X. Chen, W.~Chen, T.~G. Steele, Y.~Zhang, D.~Zhou, {Charmoniumlike states with the exotic quantum number JPC=3{\ensuremath{-}}+}, Phys. Rev. D 110~(9) (2024) 094031.
\newblock \href {http://arxiv.org/abs/2408.04828} {\path{arXiv:2408.04828}}, \href {https://doi.org/10.1103/PhysRevD.110.094031} {\path{doi:10.1103/PhysRevD.110.094031}}.

\bibitem{Cui:2013xla}
C.-Y. Cui, Y.-L. Liu, M.-Q. Huang, {Could $Z_c$(4025) be a $J^P$ = $1^+ D^* \bar{D^*}$ molecular state?}, Eur. Phys. J. C 73~(12) (2013) 2661.
\newblock \href {http://arxiv.org/abs/1308.3625} {\path{arXiv:1308.3625}}, \href {https://doi.org/10.1140/epjc/s10052-013-2661-x} {\path{doi:10.1140/epjc/s10052-013-2661-x}}.

\bibitem{Yang:1993bp}
K.-C. Yang, W.~Y.~P. Hwang, E.~M. Henley, L.~S. Kisslinger, {QCD sum rules and neutron-proton mass difference}, Phys. Rev. D 47 (1993) 3001--3012.
\newblock \href {https://doi.org/10.1103/PhysRevD.47.3001} {\path{doi:10.1103/PhysRevD.47.3001}}.

\bibitem{Narison:2002hk}
S.~Narison, {Light and heavy quark masses, flavor breaking of chiral condensates, meson weak leptonic decay constants in QCD} (2 2002).
\newblock \href {http://arxiv.org/abs/hep-ph/0202200} {\path{arXiv:hep-ph/0202200}}.

\bibitem{Gimenez:2005nt}
V.~Gimenez, V.~Lubicz, F.~Mescia, V.~Porretti, J.~Reyes, {Operator product expansion and quark condensate from lattice QCD in coordinate space}, Eur. Phys. J. C 41 (2005) 535--544.
\newblock \href {http://arxiv.org/abs/hep-lat/0503001} {\path{arXiv:hep-lat/0503001}}, \href {https://doi.org/10.1140/epjc/s2005-02250-9} {\path{doi:10.1140/epjc/s2005-02250-9}}.

\bibitem{Jamin:2002ev}
M.~Jamin, {Flavor-symmetry breaking of the quark condensate and chiral corrections to the Gell-Mann-Oakes-Renner relation}, Phys. Lett. B 538 (2002) 71--76.
\newblock \href {http://arxiv.org/abs/hep-ph/0201174} {\path{arXiv:hep-ph/0201174}}, \href {https://doi.org/10.1016/S0370-2693(02)01951-2} {\path{doi:10.1016/S0370-2693(02)01951-2}}.

\bibitem{Ioffe:2002be}
B.~L. Ioffe, K.~N. Zyablyuk, {Gluon condensate in charmonium sum rules with three-loop corrections}, Eur. Phys. J. C 27 (2003) 229--241.
\newblock \href {http://arxiv.org/abs/hep-ph/0207183} {\path{arXiv:hep-ph/0207183}}, \href {https://doi.org/10.1140/epjc/s2002-01099-8} {\path{doi:10.1140/epjc/s2002-01099-8}}.

\bibitem{Ovchinnikov:1988gk}
A.~A. Ovchinnikov, A.~A. Pivovarov, {QCD sum rule calculation of the quark gluon condensate}, Sov. J. Nucl. Phys. 48 (1988) 721--723.

\bibitem{Ellis:1996xc}
J.~R. Ellis, E.~Gardi, M.~Karliner, M.~A. Samuel, {Renormalization-scheme dependence of Pad{\'e} summation in QCD}, Phys. Rev. D 54 (1996) 6986--6996.
\newblock \href {http://arxiv.org/abs/hep-ph/9607404} {\path{arXiv:hep-ph/9607404}}, \href {https://doi.org/10.1103/PhysRevD.54.6986} {\path{doi:10.1103/PhysRevD.54.6986}}.

\bibitem{Narison:2011xe}
S.~Narison, {Gluon condensates and precise $\overline{m}_{c,b}$ from QCD-moments and their ratios to order $\alpha_s^3$ and $\langle G^4 \rangle$}, Phys. Lett. B 706 (2012) 412--422.
\newblock \href {http://arxiv.org/abs/1105.2922} {\path{arXiv:1105.2922}}, \href {https://doi.org/10.1016/j.physletb.2011.11.058} {\path{doi:10.1016/j.physletb.2011.11.058}}.

\bibitem{Narison:2018dcr}
S.~Narison, {QCD parameter correlations from heavy quarkonia}, Int. J. Mod. Phys. A 33~(10) (2018) 1850045, [Addendum: Int.J.Mod.Phys.A 33, 1892004 (2018)].
\newblock \href {http://arxiv.org/abs/1801.00592} {\path{arXiv:1801.00592}}, \href {https://doi.org/10.1142/S0217751X18500458} {\path{doi:10.1142/S0217751X18500458}}.

\end{thebibliography}

\end{document}